\documentclass[10pt]{article}

\usepackage{amsmath}
\usepackage{amssymb}

\usepackage{graphicx}

\usepackage{cite}

\usepackage{color} 

\usepackage{hyperref}
\hypersetup{pdfborder={0 0 0}} 

\usepackage[labelfont=bf,labelsep=period,justification=raggedright]{caption}

\makeatletter
\renewcommand{\@biblabel}[1]{\quad#1.}
\makeatother

\date{}

\begin{document}

\begin{flushleft}
{\Large
\textbf{Impact of adaptation currents on synchronization of coupled exponential integrate-and-fire neurons}
}
\\
Josef Ladenbauer$^{1,2\ast}$, 
Moritz Augustin$^{1}$,
LieJune Shiau$^{3}$, 
Klaus Obermayer$^{1,2}$
\\
\bf{1} Department of Software Engineering and Theoretical Computer Science, Technische Universit\"at Berlin, Berlin, Germany
\\
\bf{2} Bernstein Center for Computational Neuroscience Berlin, Berlin, Germany
\\
\bf{3} Department of Mathematics, University of Houston, Houston, Texas, United States of America
\\
$\ast$ E-mail: jl@ni.tu-berlin.de
\end{flushleft}

\section*{Abstract}

The ability of spiking neurons to synchronize their activity in a network depends on the response behavior of these neurons as quantified by the phase response curve (PRC) and on coupling properties. The PRC characterizes the effects of transient inputs on spike timing and can be measured experimentally.
Here we use the adaptive exponential integrate-and-fire (aEIF) neuron model to determine how subthreshold and spike-triggered slow adaptation currents shape the PRC. Based on that, we predict how synchrony and phase locked states of coupled neurons change in presence of synaptic delays and unequal coupling strengths.
We find that increased subthreshold adaptation currents cause a transition of the PRC from only phase advances to phase advances and delays in response to excitatory perturbations. Increased spike-triggered adaptation currents on the other hand predominantly skew the PRC to the right. Both adaptation induced changes of the PRC are modulated by spike frequency, being more prominent at lower frequencies.
Applying phase reduction theory, we show that subthreshold adaptation stabilizes synchrony for pairs of coupled excitatory neurons, while spike-triggered adaptation causes locking with a small phase difference, as long as synaptic heterogeneities are negligible. For inhibitory pairs synchrony is stable and robust against conduction delays, and adaptation can mediate bistability of in-phase and anti-phase locking.
We further demonstrate that stable synchrony and bistable in/anti-phase locking of pairs carry over to synchronization and clustering of larger networks.
The effects of adaptation in aEIF neurons on PRCs and network dynamics qualitatively reflect those of biophysical adaptation currents in detailed Hodgkin-Huxley-based neurons, which underscores the utility of the aEIF model for investigating the dynamical behavior of networks. Our results suggest neuronal spike frequency adaptation as a mechanism synchronizing low frequency oscillations in local excitatory networks, but indicate that inhibition rather than excitation generates coherent rhythms at higher frequencies.

\section*{Author Summary}

Synchronization of neuronal spiking in the brain is related to cognitive functions, such as perception, attention, and memory. It is therefore important to determine which properties of neurons influence their collective behavior in a network and to understand how. A prominent feature of many cortical neurons is spike frequency adaptation, which is caused by slow transmembrane currents. We investigated how these adaptation currents affect the synchronization tendency of coupled model neurons. Using the efficient adaptive exponential integrate-and-fire (aEIF) model and a biophysically detailed neuron model for validation, we found that increased adaptation currents promote synchronization of coupled excitatory neurons at lower spike frequencies, as long as the conduction delays between the neurons are negligible. Inhibitory neurons on the other hand synchronize in presence of conduction delays, with or without adaptation currents. Our results emphasize the utility of the aEIF model for computational studies of neuronal network dynamics. We conclude that adaptation currents provide a mechanism to generate low frequency oscillations in local populations of excitatory neurons, while faster rhythms seem to be caused by inhibition rather than excitation.

\section*{Introduction}

Synchronized oscillating neural activity has been shown to be 
involved in a variety of cognitive functions 
\cite{Singer1999, Wang2010} such as multisensory integration 
\cite{Roelfsema1997, Ghazanfar2008}, conscious perception 
\cite{Melloni2007, Hipp2011}, selective attention \cite{Fries2001, Doesburg2008}
and memory \cite{Herrmann2004, Lengyel2005}, as well as in 
pathological states including Parkinson's disease \cite{Hammond2007}, 
schizophrenia \cite{Uhlhaas2010}, and epilepsy \cite{Zijlmans2009}. 
These observations have led to a great interest in understanding the mechanisms of neuronal synchronization, how synchronous oscillations are initiated, maintained, and destabilized.
\newline

\noindent The phase response curve (PRC) provides a powerful tool to 
study neuronal synchronization \cite{Smeal2010}. The PRC is an experimentally obtainable measure that characterizes the effects of transient inputs to a periodically spiking neuron on the timing 
of its subsequent spike.
PRC based techniques have been applied widely 
to analyze rhythms of neuronal populations and have yielded valuable 
insights into, for example, motor pattern generation \cite{Ermentrout1994}, 
the hippocampal theta rhythm \cite{Netoff2005}, and memory retrieval 
\cite{Lengyel2005}. The shape of the PRC is strongly affected by ionic 
currents that mediate spike frequency adaptation (SFA) 
\cite{Crook1998, Ermentrout2012}, a prominent feature of neuronal dynamics shown by 
a decrease in instantaneous spike rate during a sustained current injection 
\cite{McCormick1985, Connors1990, LaCamera2006}. These 
adaptation currents modify the PRC in distinct ways, depending on whether 
they operate near rest or during the spike \cite{Ermentrout2012}. Using 
biophysical neuron models, it has been shown that a low threshold 
outward current, such as the muscarinic voltage-dependent $\mathrm K^+$-current 
($I_m$), can produce a \mbox{type II} PRC, characterized by phase advances and 
delays in response to excitatory stimuli, in contrast to only phase advances, defining a \mbox{type I} PRC. A high threshold outward current on the other hand, such as 
the $\mathrm Ca^{2+}$-dependent afterhyperpolarization 
$\mathrm K^+$-current ($I_{ahp}$), flattens the PRC at early phases and skews 
its peak towards the end of the period \cite{Ermentrout2001, Stiefel2009, Ermentrout2012}. Both changes of the PRC indicate an 
increased propensity for synchronization of coupled excitatory cells 
\cite{Ermentrout2001}, and can be controlled selectively through cholinergic 
neuromodulation. In particular, $I_m$ and $I_{ahp}$ are reduced by
acetylcholine with different sensitivities, which modifies the PRC shape 
\cite{Madison1987, Stiefel2008, Stiefel2009}.
\newline

\noindent In recent years substantial efforts have been exerted to develop 
single neuron models of reduced complexity that can reproduce a large 
repertoire of observed neuronal behavior, while being computationally less 
demanding and, more importantly, easier to understand and analyze than 
detailed biophysical models. Two-dimensional variants of the leaky 
integrate-and-fire neuron model have been proposed which take into 
consideration an adaptation mechanism that is spike triggered 
\cite{Treves1993} or subthreshold, capturing resonance properties 
\cite{Richardson2003}, as well as an improved description of spike initiation 
by an exponential term \cite{Fourcaud2003}. A popular example is the adaptive exponential leaky integrate-and-fire (aEIF) model by Brette and Gerstner \cite{Brette2005, Gerstner2009}. The aEIF 
model is similar to the two-variable model of Izhikevich 
\cite{Izhikevich2003}, such that both models include a sub-threshold as well as a 
spike-triggered adaptation component in one adaptation current. The advantages of the aEIF model, as opposed to the Izhikevich model, are the exponential description of spike initiation instead of a quadratic nonlinearity, and more importantly, that its parameters are of physiological relevance. Despite their 
simplicity, these two models (aEIF and Izhikevich) can capture a 
broad range of neuronal dynamics 
\cite{Izhikevich2004, Touboul2008, Naud2008} which renders them appropriate for application in large-scale 
network models \cite{Izhikevich2008, Destexhe2009}. Furthermore, the aEIF 
model has been successfully fit to Hodgkin-Huxley-type neurons as well as to
recordings from cortical neurons \cite{Brette2005, Clopath2007, Jolivet2008}.
Since lately, this model is also implemented in neuromorphic hardware systems \cite{Bruederle2011}.
\newline

\noindent Because of subthreshold and spike-triggered contributions to the adaptation current, the aEIF model exhibits a rich dynamical structure \cite{Touboul2008}, and can be tuned to reproduce the behavior of all major classes of neurons,
as defined electrophysiologically in vitro \cite{Naud2008}. 
Here, we use the aEIF model to study the influence of adaptation on network dynamics, particularly synchronization and phase locking, taking into account conduction delays and unequal synaptic strengths. 
First, we show how both subthreshold and spike-triggered adaptation affect the PRC as a function of spike frequency. Then, we apply phase 
reduction theory, assuming weak coupling, to explain how 
the changes in phase response behavior determine phase locking of neuronal 
pairs, considering conduction delays and heterogeneous synaptic strengths. 
We next present numerical simulations of networks which support the findings from
our analysis of phase locking in neuronal pairs, and show their robustness 
against heterogeneities. Finally, to validate the biophysical implication of the adaptation parameters in the aEIF model, we relate and compare the results using this model to the effects of 
$I_m$ and $I_{ahp}$ on synchronization in Hodgkin-Huxley-type conductance 
based neurons. Thereby, we demonstrate that the basic description of an adaptation current in the low-dimensional aEIF model suffices to capture the  characteristic changes of PRCs, and consequently the effects on phase locking and network behavior, mediated by biophysical adaptation currents in a complex neuron model. The aEIF model thus represents a useful and efficient tool to examine the dynamical behavior of neuronal networks.

\section*{Methods}

\subsection*{aEIF neuron model}

\noindent The aEIF model consists of two differential equations and a 
reset condition,
\begin{align} \label{eq_adex_membrane_current}
C \frac{dV}{dt} & = -g_L (V-E_L) + g_L \, \Delta_T \, e^{\tfrac{V-V_T}{\Delta_T}} -w +I \\
\tau_w \frac{dw}{dt} & = a(V-E_L) -w \label{eq_adex_adaptation_current}
\end{align}
\begin{equation} \label{eq_adex_reset}
\mathrm{if}\ V \geq V_{cut}\ \mathrm{then}\ \begin{cases}
                                                V=V_r \\ 
                                                w=w+b.
                                         \end{cases}
\end{equation}
The first equation \eqref{eq_adex_membrane_current} is the membrane equation, 
where the 
capacitive current through the membrane with capacitance $C$ 
equals the sum of ionic currents, the adaptation current $w$, and 
the input current $I$. The ionic currents are given by an ohmic 
leak current, determined by the leak conductance $g_L$ and the leak 
reversal potential $E_L$, and a $\mathrm{Na}^+$-current which is responsible for 
the generation of spikes. The $\mathrm{Na}^+$-current is approximated by the exponential 
term, where $\Delta_T$ is the threshold slope factor and $V_T$ is the 
threshold potential, assuming that the activation of 
$\mathrm{Na}^{+}$-channels is instantaneous and neglecting their 
inactivation \cite{Fourcaud2003}. The membrane time constant is 
$\tau_m:=C/g_L$. When $I$ drives the membrane potential $V$ beyond 
$V_T$, the exponential term actuates a positive feedback and leads 
to a spike, which is said to occur at the time when $V$ diverges 
towards infinity. In practice, integration of the model equations 
is stopped when $V$ reaches a finite ``cutoff'' value $V_{cut}$, 
and $V$ is reset to $V_r$ \eqref{eq_adex_reset}. 
Equation \eqref{eq_adex_adaptation_current} governs the dynamics 
of $w$, with the adaptation time constant $\tau_w$. $a$ quantifies a 
conductance that mediates subthreshold adaptation. Spike-triggered 
adaptation is included through the increment $b$ \eqref{eq_adex_reset}. 
\newline

\noindent The dynamics of the model relevant to our study is outlined as follows. 
When the input current $I$ to the neuron at rest is slowly 
increased, at some critical current the resting state is destabilized 
which leads to repetitive spiking for large regions in parameter space \cite{Naud2008}. This onset of spiking corresponds 
to a saddle-node (SN) bifurcation if $a \tau_w < g_L \tau_m$, and a 
subcritical Andronov-Hopf (AH) bifurcation if $a \tau_w > g_L \tau_m$ 
at current values $I_{SN}$ and $I_{AH}$ respectively which can be calculated 
explicitly \cite{Touboul2008}. In the former case a stable fixed 
point (the neuronal resting state) and an unstable fixed point (the saddle) merge and 
disappear, in the latter case the stable fixed point becomes unstable 
before merging with the saddle. In the limiting case 
$a \tau_w=g_L \tau_m$, both bifurcations (SN and AH) meet and the 
system undergoes a Bogdanov-Takens (BT) bifurcation. 
The sets of points with $dV/dt=0$ 
and $dw/dt=0$ are called $V$-nullcline and $w$-nullcline, respectively. It is obvious that all fixed points in the two-dimensional state space can be identified as intersections 
of these two nullclines. Spiking can occur at a 
constant input current lower than $I_{SN}$ or $I_{AH}$ depending on 
whether the sequence of reset points lies exterior to the basin of 
attraction of the stable fixed point. This means, the system just 
below the bifurcation current can be bistable; periodic spiking and constant membrane potential are 
possible at the same input current. Thus, periodic spiking trajectories do not necessarily emerge from a SN or AH bifurcation. 
We determined the lowest input 
current that produces repetitive spiking (the rheobase current, $I_{rh}$) 
numerically by delivering long-lasting rectangular current pulses to 
the model neurons at rest. Note that in general $I_{rh}$ depends on $V_r$, such that
in case of bistability, $I_{rh}$ can be reduced by decreasing $V_r$ \cite{Touboul2008}.
\newline

\noindent We selected realistic values for the model parameters ($C=0.1~\mathrm{nF}$, 
$g_L=0.01~\mathrm{\mu S}$, $E_L=-70~\mathrm{mV}$, $\Delta_T=2~\mathrm{mV}$, $V_T=-50~\mathrm{mV}$, 
$\tau_w=100~\mathrm{ms}$, $V_r=-60~\mathrm{mV}$) and varied the adaptation parameters within reasonable ranges 
($a \in [0, 0.1]~\mathrm{\mu S}$, $b \in [0, 0.2]~\mathrm{nA}$). All model parametrizations in this study lead to periodic spiking for sufficiently large $I$, possibly including transient adaptation. Parameter regions which lead to bursting and irregular spiking \cite{Naud2008} are not considered in this study.
$V_{cut}$ was set to $-30~\mathrm{mV}$, since from this value, even without an 
input current, $V$ would rise to a typical peak value of the action 
potential ($<50~\mathrm{mV}$) within less than $1~\mathrm{\mu s}$ while $w$ essentially 
does not change due to its large time constant. Only in Fig.~\ref{fig_1}A-C we used
$V_{cut} = 20~\mathrm{mV}$ to demonstrate the steep increase of $V$ past $V_T$. 

\subsection*{Traub neuron model}

\noindent In order to compare the effects of adaptation in the aEIF model 
with those of $I_m$ and $I_{ahp}$ in a biophysically detailed model and  
with previously published results 
\cite{Ermentrout2001, Jeong2007, Ermentrout2012} we used a variant of the 
conductance based neuron model described by Traub et al. \cite{Traub1991}.
The current-balance equation of this model is given by
\begin{align} \label{eq_traub_currentbalance}
C \frac{dV}{dt} & = I-I_L-I_{Na}-I_K-I_{Ca}-I_m-I_{ahp},
\end{align}
where the ionic currents consist of a 
leak current $I_L=g_L (V-E_L )$, a $\mathrm {Na}^+$-current 
$I_{Na}=g_{Na} m^3 h(V-E_{Na} )$, a delayed rectifying 
$\mathrm{K}^+$-current $I_K=g_K n^4 (V-E_K )$, a high-threshold 
$\mathrm{Ca}^{2+}$-current $I_{Ca}=g_{Ca} m_\infty (V-E_{Ca} )$ with 
$m_\infty=1/(1+exp(-(V+25)/2.5))$, and the slow $\mathrm{K}^+$-currents
$I_m=g_m \omega(V-E_K )$, and 
$I_{ahp}=g_{ahp} ([Ca^{2+}]/([Ca^{2+}]+1))(V-E_K )$. The gating variables 
$m$, $h$ and $n$ satisfy first-order kinetics 
\begin{align} 
\frac{dm}{dt} & =  \alpha_m (1-m) - \beta_m m \label{eq_traub_m} \\
\frac{dh}{dt} & =  \alpha_h (1-h) - \beta_h h \label{eq_traub_h} \\
\frac{dn}{dt} & =  \alpha_n (1-n) - \beta_n n, \label{eq_traub_n}
\end{align}
with 
$\alpha_m=0.32(V+54)/(1-exp(-(V+54)/4))$ and 
$\beta_m=0.28(V+27)/(exp((V+27)/5)-1)$,
$\alpha_h=0.128exp(-(V+50)/18)$ and 
$\beta_h=4/(1+exp(-(V+27)/5))$,
$\alpha_n=0.032(V+52)/(1-exp(-(V+52)/5))$ and 
$\beta_n=0.5exp(-(V+57)/40)$. 
The fraction $\omega$ of open $\mathrm{K}^+$-channels is governed by 
\begin{align} \label{eq_traub_omega}
\frac{d\omega}{dt} & =  \frac{\omega_\infty - \omega}{\tau_\omega}, 
\end{align}
where 
$\omega_\infty=1/(1+exp(-(V+35)/10))$, 
$\tau_\omega=100/(3.3 exp((V+35)/20)+exp(-(V+35)/20))$, and the 
intracellular $\mathrm{Ca}^{2+}$ concentration $[Ca^{2+}]$ is described by 
\begin{align}
\frac{d[Ca^{2+}]}{dt} & =  -\gamma I_{Ca} - \frac{[Ca^{2+}]}{\tau_{Ca}}. \label{eq_traub_caconcentration}
\end{align}
Units are mV for the membrane potential and ms for time. Note that the state space of the Traub model eqs.~\eqref{eq_traub_currentbalance}--\eqref{eq_traub_caconcentration} is six-dimensional. \newline

\noindent The dynamics of interest is described below. Starting from a resting state, 
as $I$ is increased, the model goes to repetitive spiking. 
Depending on the level of $I_m$, this (rest-spiking) transition occurs through a SN 
bifurcation for low values of $I_m$ or a subcritical AH bifurcation for high values 
of $I_m$, at input currents 
$I_{SN}$ and $I_{AH}$, respectively. The SN bifurcation gives rise to a 
branch of stable periodic solutions (limit cycles) with arbitrarily low 
frequency. Larger values of $I_m$ cause the stable fixed point to lose 
its stability by an AH bifurcation (at $I_{AH} < I_{SN}$). In this case, 
a branch of unstable periodic orbits emerges, which collides with a branch of 
stable limit cycles with finite frequency in a fold limit cycle bifurcation at 
current $I_{FLC} < I_{AH}$. The branch of stable periodic spiking trajectories
extends for currents larger than $I_{AH}$ and $I_{SN}$. This means that in the 
AH bifurcation regime, the model exhibits hysteresis. That is, for an input current between $I_{FLC}$ 
and $I_{AH}$ a stable equilibrium point and a stable limit cycle coexist. 
On the contrary, $I_{ahp}$ does not affect the bifurcation of the equilibria, since it is 
essentially nonexistent at rest. 
\newline

\noindent We used parameter values as in 
\cite{Ermentrout2001}. Assuming a cell surface area of 
$0.02~\mathrm{mm}^2$, the 
membrane capacitance was $C=0.2~\mathrm{nF}$, the conductances (in 
$\mathrm{\mu S}$) were 
$g_L=0.04$, $g_{Na}=20$, $g_K=16$, $g_{Ca}=0.2$, $g_m \in [0, 0.1]$, 
$g_{ahp} \in [0, 0.2]$, and the reversal potentials (in mV) were 
$E_L=-67$, $E_{Na}=50$, $E_K=-100$, $E_{Ca}=120$; 
$\gamma = 0.01$~$\mathrm{\mu M}~(\textrm{ms nA})^{-1}$ and $\tau_{Ca} = 80~\mathrm{ms}$.

\subsection*{Network simulations}

We considered networks of $N$ coupled neurons with identical 
properties using both models (aEIF and Traub), driven to repetitive spiking with period $T$,
\begin{equation} \label{eq_network_neuroncoupling}
\frac{d \mathbf x_i}{dt}=\mathbf{f}(\mathbf x_i)+ 
\sum_{j=1}^N \mathbf h_{ij} (\mathbf x_i,\mathbf x_j),
\end{equation}
where the vector $\mathbf x_i$ consists of the state variables of 
neuron $i$ 
($\mathbf x_i= (V_i,w_i)^T$ for the aEIF model, 
or $\mathbf x_i=(V_i,m_i,h_i,n_i,\omega_i,[Ca]_i)^T$ for the Traub model), 
$\mathbf{f}$ governs the 
dynamics of the uncoupled neuron (according to either neuron model) 
and the coupling function $\mathbf h_{ij}$  contains 
the synaptic current $I_{syn}$ (received by postsynaptic neuron $i$ 
from presynaptic neuron $j$) in the first component and all other 
components are zero. $I_{syn}$ was modeled using a bi-exponential 
description of the synaptic conductance,
\begin{align} 
I_{syn} (V_i,V_j ) & =g_{ij} \, s(t-d_{ij} )
        (E_{syn}-V_i ) \label{eq_biexpsynapse_current} \\
s(t) & = c \displaystyle \sum_{t_j\le t}
        \left ( e^{-\tfrac{t-t_j}{\tau_d}}-
        e^{-\tfrac{t-t_j}{\tau_r}} \right ), \label{eq_biexpsynapse_openprob}
\end{align}
where $g_{ij}$ denotes the peak conductance, $s$ the fraction of open 
ion channels, $d_{ij}$ the conduction delay which includes axonal as well as 
dendritic contributions, and $E_{syn}$ the synaptic reversal potential.
$c$ is a normalization factor which was chosen such that the peak of $s$ equals one. 
The spike times $t_j$ of neuron $j$ (at the soma) correspond to the times 
at which the membrane potential reaches $V_{cut}$ (in the aEIF model) or 
the peak of the action potential (in the Traub model).
$\tau_r$ and $\tau_d$ are the rise and decay time constants, respectively. For 
excitatory synapses the parameters were chosen to model an AMPA-mediated 
current ($E_{syn}=0~\mathrm{mV}$, $\tau_r=0.1~\mathrm{ms}$, $\tau_d=1~\mathrm{ms}$), the 
parameters for inhbitory synapses we set to describe a $\mathrm{GABA_A}$-mediated 
current ($E_{syn}=-80~\mathrm{mV}$, $\tau_r = 0.5~\mathrm{ms}$, $\tau_d=5~\mathrm{ms}$).
\newline

\noindent We simulated the aEIF and Traub neuron networks, respectively, 
taking $F := T^{-1} = 40~\mathrm{Hz}$, homogeneous all-to-all connectivity without self-feedback 
($g_{ii}=0$), and neglecting conduction delays ($d_{ij}=0$). We further
introduced heterogeneities of several degrees w.r.t. synaptic strengths and 
conduction delays to the computationally less demanding aEIF network. 
Specifically, $g_{ij}$ ($i \ne j$) and $d_{ij}$ were 
sampled from a uniform distribution over various value ranges. 
The neurons were weakly coupled, in the sense that the total synaptic input received by a neuron from all other neurons in the network (assuming they spike synchronously) resulted in a maximal change of ISI ($T$) of less than 5\%, which was determined by simulations.
As initial 
conditions we used points of the spiking trajectory at times that were uniformly sampled from the interval $[0,T]$, i.e. the initial states were asynchronous. 
Simulation time was $20~\mathrm{s}$ for each configuration of the aEIF networks and $10~\mathrm{s}$ 
for the Traub neuron networks. All network simulations were done with BRIAN 1.3 
\cite{Goodman2008} applying the second-order 
Runge-Kutta integration method with a time step of $1~\mathrm{\mu s}$ for coupled pairs and $10~\mathrm{\mu s}$ for larger networks.
\newline

\noindent We measured the degree of spike synchronization in the simulated 
networks using averaged pairwise cross-correlations between the neurons 
\cite{Wang1996},
\begin{equation} \label{eq_coherence_measure_wang}
\kappa = \left \langle \frac{\sum_k s_i^k s_j^k}
        {\sqrt{ \sum_k s_i^k \sum_k s_j^k }}  \right \rangle,
\end{equation}
where $s_i^k=1$ if neuron $i$ spikes in time interval $k$, otherwise 
$s_i^k=0$, for $k=1,\dots,T_\kappa/\tau$. $\langle . \rangle$ indicates 
the average over all neuronal pairs ($i,j$) in the network. 
Calculation period $T_\kappa$ was $1~\mathrm{s}$ and time bin $\tau$ was $2.5~\mathrm{ms}$. 
$\kappa$ assumes a value of $0$ for asynchronous spiking and approaches $1$ for perfect 
synchronization.
\newline

\noindent In order to quantify the degree of phase locking of neurons in the network 
we applied the mean phase coherence measure $\sigma$ 
\cite{Mormann2000, Fink2011} defined by
\begin{equation} \label{eq_locking_measure_fink}
\sigma = \left \langle \left | 
        \frac{1}{K} \sum_{k=1}^K e^{i \varphi^k_{ij}} 
        \right | \right \rangle,
\end{equation}
where $\varphi_{ij}^k$ is the phase difference between neurons $i$ and $j$ 
at the time of the $k^{th}$ spike $t_i^k$ of neuron $i$, 
$\varphi_{ij}^k=2 \pi(t_i^k-t_j^k)/(t_j^{k+1}-t_j^k)$. $t_j^k$ is the largest 
spike time of neuron $j$ that precedes $t_i^k$, $t_j^{k+1}$ is the smallest 
spike time of neuron $j$ that succeeds $t_i^k$. $K$ is the number of 
spikes of neuron $i$ in the calculation period 
$T_K$. $| \sum_l e^{i\varphi_l} | = 
  \sqrt{(\sum_l \cos \varphi_l)^2 + (\small \sum_l \sin \varphi_l)^2}$ 
 and 
$\langle . \rangle$ denotes the average over all pairs $(i,j)$.
$\sigma=0$ means no neuronal pair phase locks, $\sigma=1$ indicates complete 
phase locking. $\sigma$ was calculated using for $T_K$ the last $10~\mathrm{s}$ (aEIF networks)
or $5~\mathrm{s}$ (Traub networks) of each simulation.

\subsection*{PRC calculation}

The PRC can be obtained (experimentally or in simulations) by delivering small 
perturbations to the membrane 
potential of a neuron oscillating with period $T$ at different phases $\vartheta$ 
and calculating the change of the period.
The change of period can be measured within the current cycle or several cycles after the perturbation to exclude transients which yields an asymptotic type of PRC considered here. 
The asymptotic PRC\footnote{In the following we omit the term ``asymptotic'' and just call it PRC.} is then expressed as a 
function of phase as 
$\textrm{PRC}(\vartheta) = T-T_{pert} (\vartheta)$, where  
$T_{pert} (\vartheta)$ is the period of the neuron perturbed at $\vartheta$, measured several cycles after the perturbation.
Positive 
(negative) values of $\textrm{PRC}(\vartheta)$ represent phase advances (delays).
An alternative technique of determining the PRC is to solve the linearized adjoint equation 
\cite{Ermentrout1996, Ermentrout2001, Brown2004, Ermentrout2010, Schwemmer2012}
\begin{equation} \label{eq_linear_adjoint}
\frac{d \mathbf q}{dt} = -D_{ \mathbf{x} } 
        \mathbf{f}(\mathbf{\bar x}(t))^T \mathbf q,
\end{equation}
subject to the normalization condition 
$\mathbf q(0)^T \mathbf{f}(\mathbf{\bar x}(0))=1$ (see Text S1 A). $\mathbf x$, $\mathbf{f}$ 
are as described above (cf. eq.~\eqref{eq_network_neuroncoupling}) and $D_{ \mathbf{x} } \mathbf{f}$ is the Jacobian matrix of $\mathbf{f}$. $\mathbf{\bar x}$ denotes the asymptotically stable $T$-periodic spiking trajectory as a solution of the system  
\begin{equation} \label{eq_uncoupled_system}
\frac{d \mathbf x}{dt} = \mathbf{f}(\mathbf{x}),
\end{equation}
of differential equations and a reset condition in case of the aEIF model. 
Eq.~\eqref{eq_uncoupled_system} together with the reset condition describe the dynamics of an uncoupled neuron. $\mathbf{\bar x}$ is an 
attractor of this dynamical system and nearby trajectories will converge to it. 
To obtain $\mathbf{\bar x}$, we integrated the neuron model equations for a given set of parameters and adjusted the input current $I$, such that the period was $T$.
Analysis was restricted to the regular spiking regime (cf. \cite{Naud2008} for the aEIF model). Parameter regions where bursting and chaotic spiking occurs were avoided.
\newline

\noindent For Traub model trajectories, the peak of the action potential is identified with phase $\vartheta = 0$, for aEIF trajectories $\vartheta = 0$ corresponds to the point of reset.
The first component $q^V$ of the normalized $T$-periodic solution $\mathbf q$ of eq.~\eqref{eq_linear_adjoint} represents the PRC, also called infinitesimal PRC, which characterizes the response of the oscillator to a vanishingly small perturbation (cf. Text S1 A).
For continuous limit cycles $\mathbf{\bar x}$, as produced by the Traub model, $\mathbf q$ can be obtained by solving eq.~\eqref{eq_linear_adjoint} backward in time over several periods with arbitrary initial conditions. Since $\mathbf{\bar x}$ is asymptotically stable, the $T$-periodic solution of the adjoint system, eq.~\eqref{eq_linear_adjoint}, is unstable. Thus, backward integration damps out the transients and we arrive at the periodic solution of eq.~\eqref{eq_linear_adjoint}
\cite{Hoppensteadt1997, Ermentrout2010, Schwemmer2012}.
In case of the aEIF model with an asymptotically stable $T$-periodic solution $\mathbf{\bar x}$, that involves a discontinuity in both variables $\bar V(t)$, $\bar w(t)$ at integer multiples of $T$, 
we treated the adjoint equations as a boundary value problem \cite{Ermentrout2012}. Specifically, we solved the adjoint system
\begin{align} \label{eq_adex_adjoint}
\frac{d q^V}{dt} & = \frac{g_L}{C} \left( 1 - e^{\tfrac{{\bar V}(t)-V_T}{\Delta_T}} \right) q^V - \frac{a}{\tau_w} q^w \\
\frac{d q^w}{dt} & = \frac{q^V}{C} + \frac{q^w}{\tau_w},
\end{align}
subject to the conditions
\begin{align} \label{eq_adex_adjoint_cond1}
& q^V(0) \frac{d \bar{V}}{dt}(0) + q^w(0) \frac{d \bar{w}}{dt}(0) = 1 \\
& q^w(0) = q^w(T^-), \label{eq_adex_adjoint_cond2}
\end{align}
where $q^V, q^w$ denote the two components of $\mathbf q$, and $q^w(T^-) := \mathrm{lim}_{t \nearrow T} \, q^w(t)$ is the left-sided limit.
Eq.~\eqref{eq_adex_adjoint_cond1} is the normalization condition. 
Eq.~\eqref{eq_adex_adjoint_cond2} is the continuity condition, which ensures $T$-periodicity of the solution (see Text S1 A, derivation based on \cite{Mueller1995,Samoilenko1995,Akhmet2010}).
From the fact, that the end points of $T$-periodic aEIF trajectories differ, i.e. 
$\bar V(0) = V_r$, $\bar V(T^-) = V_{cut}$ and $\bar w(0) = \bar w(T^-) + b$,
it follows that $\mathbf{f}(\mathbf{\bar x}(0)) \neq \mathbf{f}(\mathbf{\bar x}(T^-))$, which in turn leads to $\mathbf q(0) \neq \mathbf q(T^-)$. 
Perturbations of the same strength, which are 
applied to $V$ just before and after the spike, have therefore a different effect on 
the phase, leading to a discontinuity in the PRC.
\newline

\noindent
The PRCs presented in this study were calculated using the adjoint method \footnote{We solved eqs.\eqref{eq_adex_adjoint}--\eqref{eq_adex_adjoint_cond2} 
numerically using a fifth-order collocation method implemented in MATLAB.}. 
For validation purposes, we also simulated a number of PRCs by directly applying small perturbations to the membrane potential ${\bar V}$ of the oscillating neuron at different phases and measuring the change in phase after many cycles -- to ensure, that the 
perturbed trajectory had returned to the attractor $\mathbf{\bar x}$ (see Figs.~\ref{fig_2}A,B,~\ref{fig_10}C).
The results are in good agreement with the results of the adjoint method.

\subsection*{Phase reduction}

In the limit of weak synaptic interaction, which guarantees that a perturbed 
spiking trajectory remains close to the attracting (unperturbed) trajectory $\mathbf{\bar x}$, we can reduce the network model \eqref{eq_network_neuroncoupling} 
to a lower dimensional network model where neuron $i$ is described by its 
phase $\vartheta_i$
\cite{Ermentrout2010, Schwemmer2012, Hoppensteadt1997, Ermentrout1991, Kuramoto1984} as follows.
\begin{align} \label{eq_phasenet_integral}
\frac{d \vartheta_i}{dt} & = 1 + \sum_{j=1}^N \frac{1}{T} \int_0^T 
  q_i^V (s) I_{syn} (\bar V_i (s),\bar V_j (s+\vartheta_j-\vartheta_i)) ds \\
 & =: 1 + \sum_{j=1}^N H_{ij}^d(\vartheta_j - \vartheta_i), 
 \label{eq_phasenet_interaction_function}
\end{align}
where $q_i^V$ is the PRC of neuron $i$ and $\bar V_i$ the first component 
(membrane potential) of the spiking trajectory $\mathbf{\bar x}_i$ (see previous section and Text S1 B). 
$H_{ij}^d$ is the $T$-periodic averaged 
interaction function calculated using $I_{syn}$ with conduction delay $d_{ij}$ 
\eqref{eq_biexpsynapse_current}. Note that $d_{ij}$ simply causes a shift in the 
interaction function: 
$H_{ij}^d (\vartheta_j-\vartheta_i )=H_{ij}^0 (\vartheta_j-\vartheta_i-d_{ij})$. 
$H_{ij}^d$ only depends on the difference of the phases (in the argument) which 
is a useful property when analyzing the stability of phase locked states of 
coupled neuronal pairs. In this case (without self-feedback as already assumed) the phase difference 
$\varphi := \vartheta_2-\vartheta_1$
evolves according to the scalar differential equation
\begin{equation} \label{eq_scalar_phase_ode}
\frac{d\varphi}{dt}=H_{21}^d (-\varphi)-H_{12}^d (\varphi) \:
 =: H_{\Delta}(\varphi),
\end{equation}
whose stable fixed points are given by the zero crossings 
$\hat \varphi$ of $H_{\Delta}$ for which 
$\mathrm{lim}_{\varepsilon \searrow 0} \, d H_{\Delta}(\hat \varphi - \varepsilon) / d \varphi < 0$ and $\mathrm{lim}_{\varepsilon \searrow 0} \, d H_{\Delta}(\hat \varphi + \varepsilon) / d \varphi < 0$. If $H_{\Delta}$ is differentiable at $\hat \varphi$, these left and right sided limits are equal and represent the slope. Note however that $H_{\Delta}$ is continuous, but not necessarily differentiable due to the discontinuity of the PRC of an aEIF neuron. Therefore, the limits might not be equal in this case. The case where $H_{\Delta}$ is discontinuous at $\hat \varphi$, which can be caused by $\delta$-pulse coupling, i.e. $I_{syn}$ is replaced by a $\delta$-function, is addressed in the Results section. 
We calculated these stable fixed points, which correspond to stable 
phase locked states, for pairs of identical cells coupled with equal or
heterogeneous synaptic strengths and symmetric conduction delays, $d := d_{12}=d_{21}$, using PRCs 
derived from the aEIF and Traub neuron models, driven to $40~\mathrm{Hz}$ periodic 
spiking. Periodic spiking trajectories of both models and PRCs of Traub neurons were computed using variable order multistep integration methods, for PRCs of aEIF neurons a fifth-order collocation method was used to solve eqs.~\eqref{eq_adex_adjoint}--\eqref{eq_adex_adjoint_cond2}. These integration methods are implemented in MATLAB (2010a, The MathWorks). Bifurcation currents of the Traub model were calculated using MATCONT \cite{Dhooge2003, Govaerts2006}.

\section*{Results}

\subsection*{PRC characteristics of aEIF neurons}

\begin{figure}[!htp]
  \begin{center}
  \includegraphics[width=0.60\textwidth]{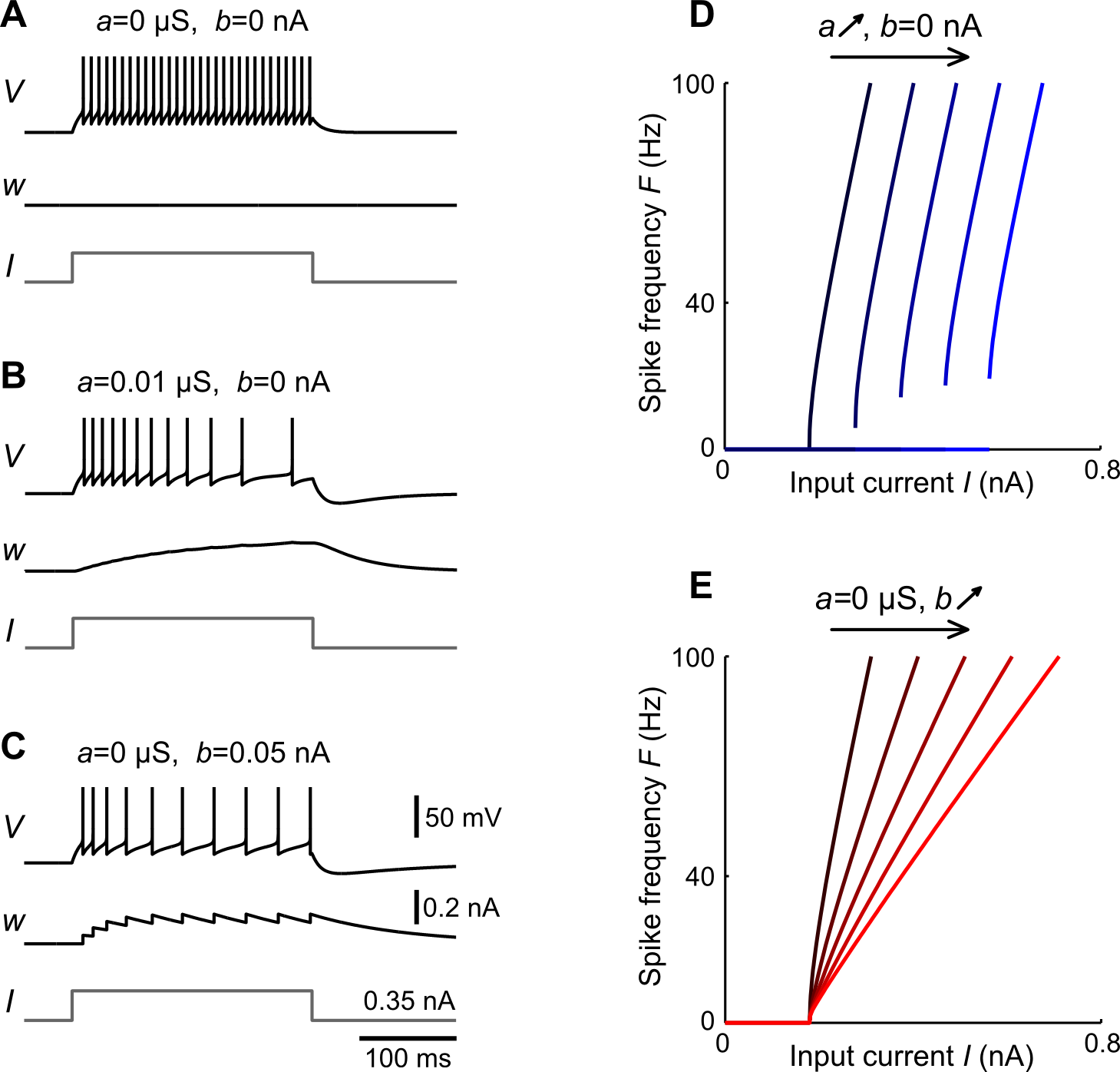}
  \end{center}
  \caption{\textbf{Influence of adaptation on spiking behavior and 
    $\mathbf F$-$\mathbf I$ curves of aEIF neurons}. 
    A-C: Membrane potential $V$ and adaptation current $w$ of aEIF neurons without adaptation 
    (A), with subthreshold adaptation (B) and with spike-triggered adaptation (C), 
    in response to step currents $I$. To demonstrate the steep increase of $V$ past $V_T$, 
    $V_{cut}$ was set to $20~\mathrm{mV}$. Note that the neuron in C has not reached its 
    steady state frequency by the end of the rectangular current pulse. D,E: \mbox{$F$-$I$} relationships for $a=0, 0.005, 0.01, 0.015, 0.02~\mathrm{\mu S}$, $b=0~\mathrm{nA}$ (black -- blue, D) and 
    $a=0~\mathrm{\mu S}$, $b=0, 0.01, 0.02, 0.03, 0.04~\mathrm{nA}$ (black -- red, E). All other model parameters used for this figure are provided in the Methods section.
  }
  \label{fig_1}
\end{figure}

We first examine the effects of the adaptation components $a$ and $b$, respectively, 
on spiking behavior of aEIF neurons at rest in response to (suprathreshold) 
current pulses (Fig.~\ref{fig_1}A-C). Without adaptation ($a=b=0$) the
model produces tonic spiking (Fig.~\ref{fig_1}A). Increasing $a$ or $b$ leads to SFA as shown 
by a gradual increase of the inter spike intervals (ISI) until a steady-state spike 
frequency $F$ is reached. Adaptation current $w$ builds up and saturates slowly when 
only conductance $a$ is considered (Fig.~\ref{fig_1}B) in comparison to spike-triggered increments $b$ (Fig.~\ref{fig_1}C). 
Fig.~\ref{fig_1}D,E depicts the relationship between $F$ and the injected current $I$ 
for various fixed values of $a$ and $b$. 
Increased subthreshold adaptation causes the minimum spike frequency to jump from zero to a positive 
value, producing a discontinuous \mbox{$F$-$I$} curve (Fig.~\ref{fig_1}D). 
A continuous (discontinuous) \mbox{$F$-$I$} curve indicates
\mbox{class I} (II) membrane excitability which is typical for a SN (AH) bifurcation at the onset of 
spiking respectively. 
An increase of $a$ causes this bifurcation to switch from SN to AH, thereby 
changing the membrane excitability from \mbox{class I} to II, shown by the \mbox{$F$-$I$} curves. An increase of $b$ on the other hand does not produce a discontinuity in the \mbox{$F$-$I$} curve, i.e. the membrane excitability remains \mbox{class I} (Fig.~\ref{fig_1}E).
Furthermore, increasing $a$ shifts the \mbox{$F$-$I$} curve to
larger current values without affecting its slope, while an increase of $b$ decreases the slope of the \mbox{$F$-$I$} curve in a divisive manner. 
When $b$ is large, the neuron is desensitized in the sense that
spike frequency is much less affected by changes in the driving input. 
\\

\begin{figure}[!htp]
  \begin{center}
  \includegraphics[width=1.0\textwidth]{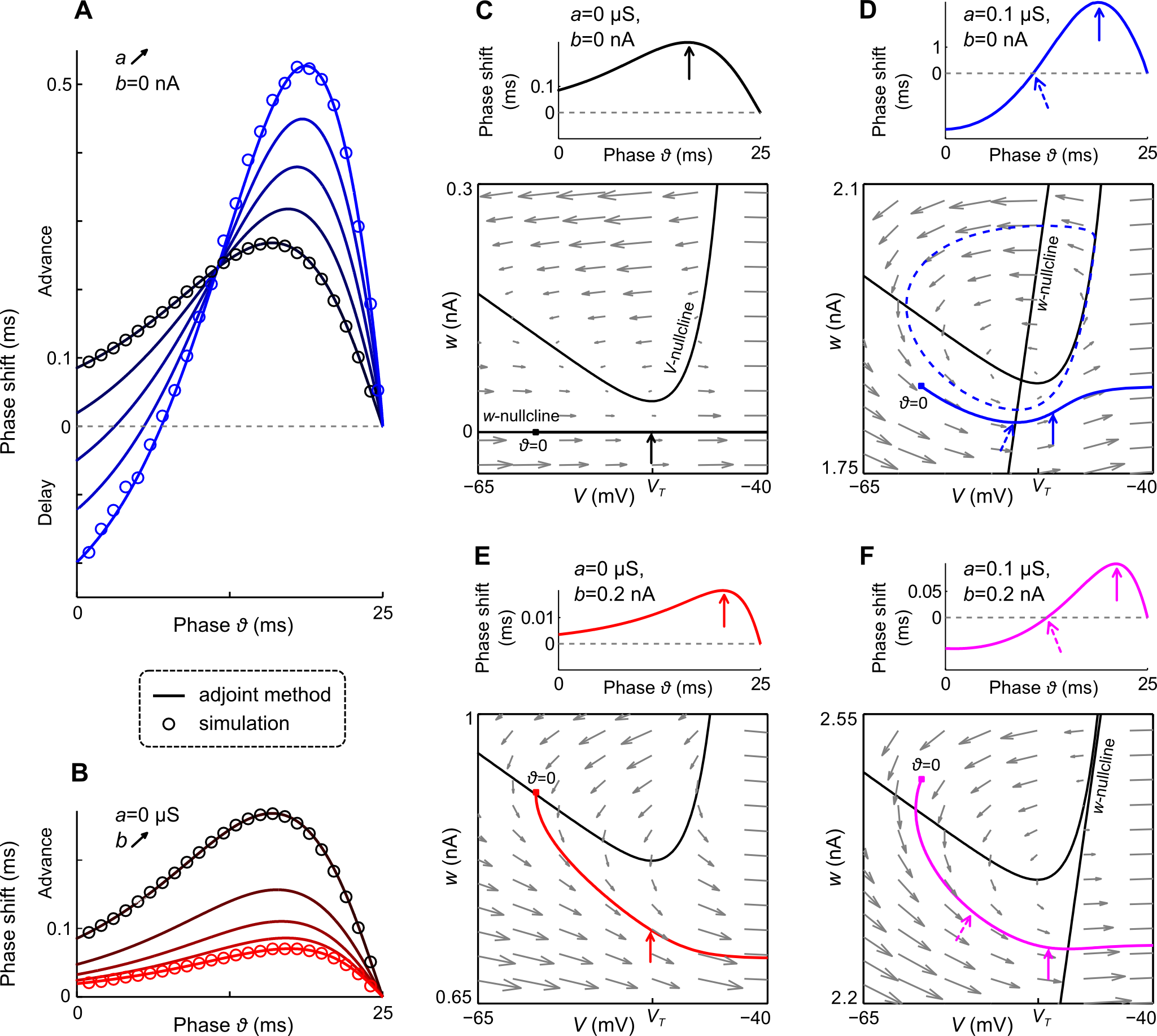}
  \end{center}
  \caption{\textbf{Effects of adaptation on PRCs of aEIF neurons}. 
    A,B: PRCs associated with adaptation parameters as in 
    Fig.~\ref{fig_1}D,E.
    Solid curves are PRCs calculated with the adjoint method and scaled by 0.1~mV, circles denote PRC
    points that were obtained from numerical simulations of eqs.~\eqref{eq_adex_membrane_current}--\eqref{eq_adex_reset}, using 0.1~mV perturbations at 
    various phases $\vartheta$ (see Methods and Text S1 A). 
    The input currents $I$ were chosen to ensure 
    40~Hz spiking. Note that the discontinuity of the PRCs at $\vartheta=0$ is 
    caused by the reset of the spiking trajectories. C-F Top: PRCs for adaptation 
    parameters as indicated and 
    $I=0.217~\mathrm{nA}$ (C), $I=2.039~\mathrm{nA}$ (D), $I=1.003~\mathrm{nA}$ (E), $I=2.530~\mathrm{nA}$ (F). 
    C-F Bottom: Vector field, $V$- and $w$-nullclines, and 
    periodic spiking trajectory in the respective state space. The reset point 
    (solid square) of
    the trajectory corresponds to the phase $\vartheta = 0$.
    A solid arrow marks the location along the trajectory where the 
    PRC (shown above) has its maximum. Dashed arrows in D, F mark the 
    trajectory points that correspond to the zero crossings of the PRCs.
    Trajectory points change slowly in regions where the vector field magnitudes are small.  
    The dashed blue curve in D 
    denotes the boundary of the domain of attraction of the fixed point, 
    which is located
    at the intersection of the nullclines. Note that differences in the vector
    fields and $V$-nullclines between C and E as well as D and F, are due to the changes in $I$.
  }
  \label{fig_2}
\end{figure}

\noindent In Fig.~\ref{fig_2}A,B we show how $a$ and $b$ differentially affect the shape of the 
PRC of an aEIF neuron driven to periodic spiking. The PRCs calculated using the adjoint method (solid curves) match well with those obtained from simulations (circles).
While non-adapting neurons have 
monophasic \mbox{(type I)} PRCs, which indicate only advancing effects of excitatory 
perturbations, increased levels of $a$ produce biphasic \mbox{(type II)} PRCs with larger magnitudes, which predict a 
delaying effect of excitatory perturbations received early in the oscillation cycle. 
An increase of $b$ on the other hand flattens the PRC at early phases, shifts its peak towards the 
end of the period and reduces its magnitude. The type of the PRC however remains
unchanged \mbox{(type I)}. Indeed, if $a=0$ the PRC must be \mbox{type I}, since in this case the component $q^V$ of the solution of the adjoint system, eqs.~\eqref{eq_adex_adjoint}--\eqref{eq_adex_adjoint_cond2}, can be written as
$q^V(t) = q^V(0) \, e^{\int_0^t{\gamma (s) ds}}$, 
where $\gamma (s)$ is given by the right-hand side of eq.~\eqref{eq_adex_adjoint}. Thus, $q^V$ cannot switch sign.  
\\

\noindent To provide an intuitive explanation for the effects of adaptation on the PRC, 
we show the vector fields, $V$- and $w$-nullclines, and periodic spiking trajectories of four aEIF neurons (Fig.~\ref{fig_2}C-F). One neuron does not have an adaptation current ($a=b=0$), two neurons possess only one adaptation mechanism ($a=0.1~\mathrm{\mu S}$, $b=0~\mathrm{nA}$ and $a=0~\mathrm{\mu S}$, $b=0.2~\mathrm{nA}$, respectively) and for one both adaptation parameters are increased ($a=0.1~\mathrm{\mu S}$, $b=0.2~\mathrm{nA}$). An excitatory perturbation to the non-adapting neuron at any point of its trajectory, i.e. at any phase, shifts this point closer to $V_{cut}$ along the trajectory, 
which means the phase is shifted closer to $T$, hence the advancing effect (Fig.~\ref{fig_2}C). The phase advance is strongest if the perturbing input is received at the position along the trajectory around which the vector field has the smallest magnitude, i.e. where the trajectory is ``slowest''.
In case of subthreshold adaptation (Fig.~\ref{fig_2}D), the adapted periodic spiking 
trajectory starts at a certain level of $w$ which decreases during the early part of the 
oscillation cycle and increases again during the late part, after the trajectory has passed 
the $w$-nullcline. A small transient excitatory input at an early phase pushes the respective point of the
trajectory to the right (along the $V$-axis) causing the perturbed trajectory to pass 
through a region above the unperturbed trajectory, somewhat closer to the fixed point around which the vector field is almost null. Consequently, 
the neuron is slowed down and the subsequent spike
delayed. An excitatory perturbation received at a later phase (to the right of the dashed arrow) causes phase advances, since the perturbed trajectory either remains nearly unchanged, however with a shorter path to the end of the cycle, compared to the unperturbed trajectory, or it passes below the unperturbed one where the magnitude of the vector field (pointing to the right) is larger. Note that for the parametrization in Fig.~\ref{fig_2}D, both, the resting state as 
well as the spiking trajectory are stable. In this case, a strong depolarizing input at an early phase can push the corresponding trajectory point into the domain of attraction of the fixed point, encircled by the dashed line in the figure, which would cause the resulting trajectory to spiral towards the fixed point and the neuron would stop spiking. 
On the other hand, increasing $I$ would shrink the domain of attraction of the fixed point and at $I = I_{AH}$, it would be destabilized by a subcritical AH bifurcation.
When $a=0$ and $b>0$, we obtain a \mbox{type I} PRC (Fig.~\ref{fig_2}E), as explained above. 
The advancing effect of an excitatory perturbation is strongest late in the oscillation cycle, indicated by the red arrow, where the perturbation pushes a trajectory point from a ``slow'' towards a ``fast'' region closer to the end of the cycle, as shown by the vector field. 
When $a$ as well as $b$ are increased, the PRC exhibits both adaptation mediated features (\mbox{type II} and skewness), see Fig.~\ref{fig_2}F. 
A push to the right along the corresponding trajectory experienced early in the cycle brings the perturbed trajectory closer to the fixed point and causes a delayed next spike. Such an effect persists even if the fixed point has disappeared due to a larger input current. In this case, the region where the fixed point used to be prior to the bifurcation, known as ``ghost'' of the fixed point, the vector field is still very small. This means that \mbox{type II} PRCs can exist for larger input currents $I > I_{SN}$. Note that differences of the vector fields and the shift of the nullclines relative to each other in Fig.~\ref{fig_2}C,D as well as Fig.~\ref{fig_2}E,F are due to different input current values (as an increase of $I$ moves the $V$-nullcline upwards). 
The maximal phase advances, 
indicated by solid arrows in Fig.~\ref{fig_2}A,B, are close to the threshold potential $V_T$ 
(where the $V$-nullcline has its minimum) in all four cases.
\\

\begin{figure}[!htp]
  \begin{center}
  \includegraphics[width=0.9\textwidth]{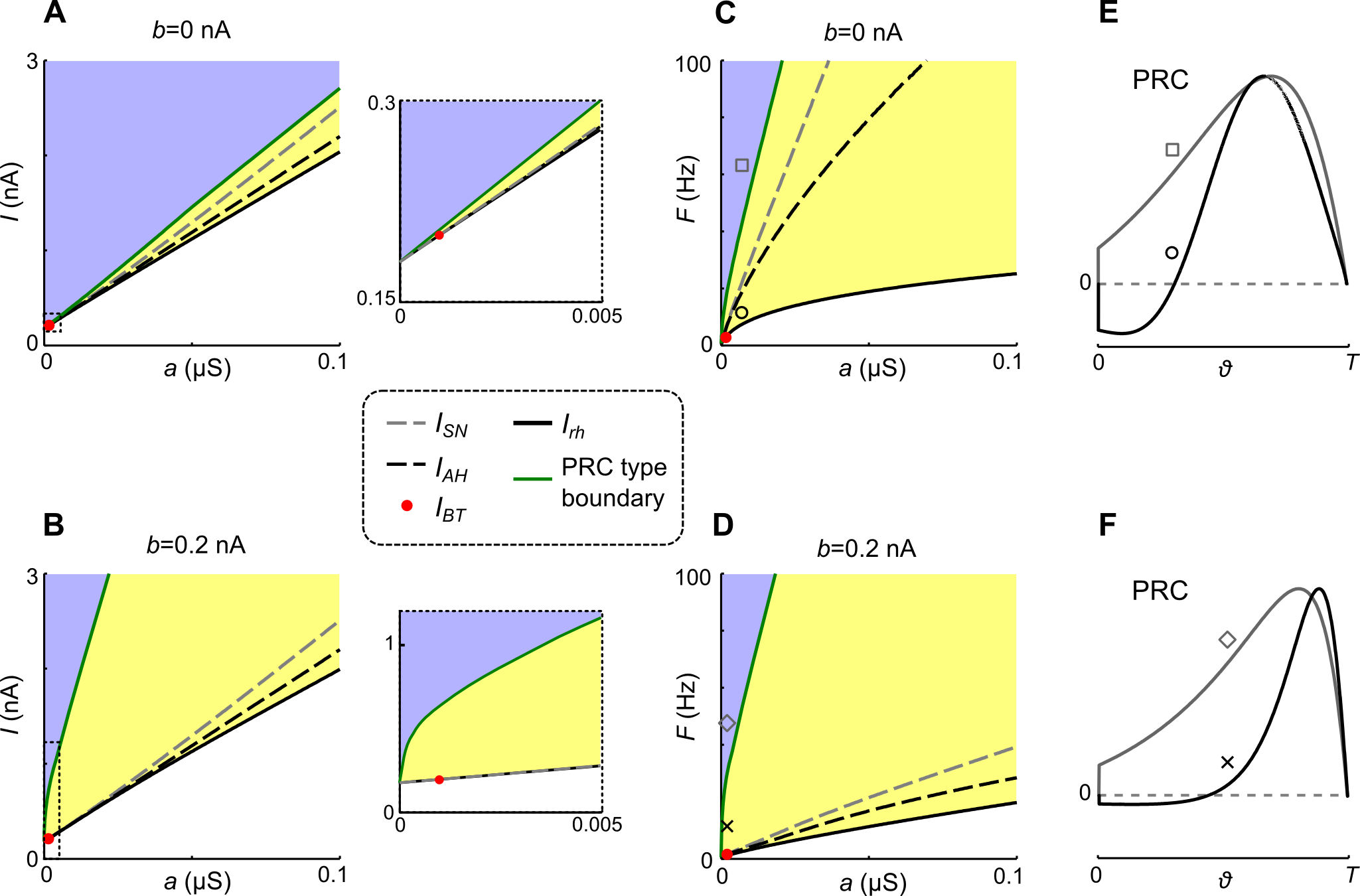}
  \end{center}  
  \caption{\textbf{Bifurcation currents of the aEIF model and dependence of PRC 
    characteristics on the input current}. 
    A,B: Rheobase current (solid black), SN and AH bifurcation currents $I_{SN}$, $I_{AH}$ 
    (dashed grey, dashed black) respectively, as well as input current (green) which separates \mbox{type I} 
    (blue) and \mbox{type II} (yellow) PRC regions, as a function of $a$, for $b=0~\mathrm{nA}$ (A) and 
    $b=0.2~\mathrm{nA}$ (B). At $a=0.001~\mathrm{\mu S}$ a BT bifurcation occurs at $I_{BT}$
    (where the SN and the AH bifurcations meet) marked by the red dot. The region around 
    $I_{BT}$ is displayed in a zoomed view. If $a<0.001~\mathrm{\mu S}$ the system 
    undergoes a SN bifurcation at $I_{SN}$, if $a>0.001~\mathrm{\mu S}$ an AH
    bifurcation occurs at $I_{AH} < I_{SN}$. C,D: Spike frequencies $F$ corresponding to the 
    input currents in A and B. Note that the region in $I$-$a$ space where the PRCs are \mbox{type II} is very shallow in A compared to B, the corresponding regions in \mbox{$F$-$a$} space shown in C and D however are rather similar. This is due to the steep (flat) \mbox{$F$-$I$} relationship for $b=0~\mathrm{nA}$ ($b=0.2~\mathrm{nA}$) respectively (see Fig.~\ref{fig_1}D,E). E,F: PRCs with locations in \mbox{$F$-$a$} space as indicated, scaled 
    to the same period $T$.
  }
  \label{fig_3}
\end{figure}

\noindent We next investigate how the changes in PRCs caused by either adaptation component 
are affected by the spike frequency. Bifurcation currents, rheobase currents and corresponding 
frequencies, in dependence of $a$ and $b$, as well as regions in parameter space where PRCs are 
\mbox{type I} and II, are displayed in Fig.~\ref{fig_3}A-D. Fig.~\ref{fig_3}E,F shows how individual
PRCs are modulated by spike frequency (input current). Both PRC characteristics, caused 
by $a$ and $b$, respectively, are more pronounced at low frequencies. Increasing $I$ 
changes a \mbox{type II} PRC to \mbox{type I} and shifts its peak towards an earlier phase. The input 
current which separates \mbox{type I} and \mbox{type II} PRC regions (in parameter space) increases with
both, $a$ and $b$ (Fig.~\ref{fig_3}A,B). That is, an increase of $b$ can also 
turn a \mbox{type I} into a \mbox{type II} PRC, by bringing the spiking trajectory closer to the fixed point or its ``ghost''. This is however only possible if the system is in the AH bifurcation regime ($a > C/ \tau_w$) or close to it. 
Spike-triggered adaptation thereby considerably influences the range of input currents for which the PRCs are \mbox{type II}.
The spike frequency according to the input current, at which a \mbox{type II} PRC turns into \mbox{type I} increases substantially with increasing $a$, but only slighly with an increase of $b$ (Fig.~\ref{fig_3}C,D). The latter can be recognized by the similarity of the respective (green) curves in the subfigures C and D. \mbox{Type II} 
PRCs thus only exist in the lower frequency band whose width increases with increasing subthreshold adaptation.

\subsection*{Phase locking of coupled aEIF pairs}

In this section, we examine how the changes in phase response properties 
due to adaptation affects phase locking of coupled pairs of periodically 
spiking aEIF neurons. Specifically, we first analyze how the shape of the PRC
determines the fixed points of eq.~\eqref{eq_scalar_phase_ode} and 
their stability, and then show how the modifications of the PRC mediated 
by the adaptation components $a$ and $b$ change those fixed points. 
Finally, we investigate the effects of conduction delays and heterogeneous 
coupling strengths on phase locking in dependence of adaptation.
\\

\noindent 
\textbf{Relation between phase locking and the PRC}

\noindent In case of identical cell pairs and symmetric synaptic strengths, 
$g_{12}=g_{21}$, the interaction functions in eq.~\eqref{eq_scalar_phase_ode} are identical, $H_{12}^d = H_{21}^d =: H^d$, where $d$ is the conduction delay. 
$H_{\Delta}(\varphi) = H^{d}(-\varphi)-H^{d}(\varphi)$ then becomes an odd, $T$-periodic function, which has roots at $\varphi=0$ and $\varphi=T/2$. Thus, the in-phase and anti-phase locked states always exist.
The stability of these two states can be ``read off'' the PRC even without having to calculate $H^d$, as is explained below. Let $\hat{\varphi} \in \{ 0,T/2 \}$ in the following. The fixed point $\hat{\varphi}$ of eq.~\eqref{eq_scalar_phase_ode} is stable if $\mathrm{lim}_{\varepsilon \searrow 0} \, d H^{d}(\hat \varphi - \varepsilon) / d \varphi > 0$ and $\mathrm{lim}_{\varepsilon \searrow 0} \, d H^{d}(\hat \varphi + \varepsilon) / d \varphi > 0$. Note that the left and right sided limits are not equal if $H^{d}$ is not differentiable at $\hat \varphi$, due to the discontinuity of the PRC of an aEIF neuron. 
\\

\noindent First, consider a synaptic current with infinitely fast rise and decay. In this case we use a positive (or negative) $\delta$-function in eq.~\eqref{eq_phasenet_integral} instead of $I_{syn}$ to describe the transient excitatory (or inhibitory) pulse.
$H^d(\varphi)$ is then given by 
\begin{equation}
\label{eq_delta_interaction}
H^d(\varphi) = 
  \pm \frac{1}{T} \int_0^T 
  q^V (s) \,\delta (s+\varphi - d \:\:\: \text{mod } T)\: ds =
  \pm \frac{1}{T} \: q^V (d - \varphi),
\end{equation} 
that is, $H^d(\varphi)$ becomes the PRC, mirrored at $\vartheta = T/2$, rightwards shifted by the delay $d$ and scaled by $\pm 1/T$. The sign of the slope of $H^{d}(\hat{\varphi})$ is thus given by the negative (positive) sign of the PRC slope at $ \vartheta = d - \hat{\varphi}$, $d \neq \hat{\varphi}$, for excitatory (inhibitory) synapses respectively. For the aEIF model, the case $d = \hat{\varphi}$ requires a distinction, because $H^{d}$ and $H_{\Delta}$ are discontinuous at $\hat \varphi$. 
Let $\Delta \varphi > 0$ be the distance between $\hat \varphi$ and the closest root of $H_{\Delta}(\varphi)$. Since $H_{\Delta}(\varphi)$ is odd and $T$-periodic, 
$H_{\Delta}(\hat{\varphi}^-) > H_{\Delta}(\hat{\varphi}^+)$ implies stability of 
$\hat \varphi$, in the sense that $\varphi$ increases on the interval $(\hat \varphi - \Delta \varphi, \hat \varphi)$ and decreases over $(\hat \varphi, \hat \varphi + \Delta \varphi)$. Thus, $\hat \varphi$ can be considered an attractor.
$H_{\Delta}(\hat{\varphi}^-) > H_{\Delta}(\hat{\varphi}^+)$ is equivalent to
$H^{d}(\hat{\varphi}^-) < H^{d}(\hat{\varphi}^+)$ which in turn is equivalent to 
$\text{PRC}( T^- ) > \text{PRC}( 0 )$ for excitatory coupling and 
$\text{PRC}( T^- ) < \text{PRC}( 0 )$ for inhibitory coupling. Hence, it is the discontinuity of the PRC which determines the stability of $\hat \varphi$ in this case.
\\

\begin{figure}[!htp]
  \begin{center}
  \includegraphics[width=0.5\textwidth]{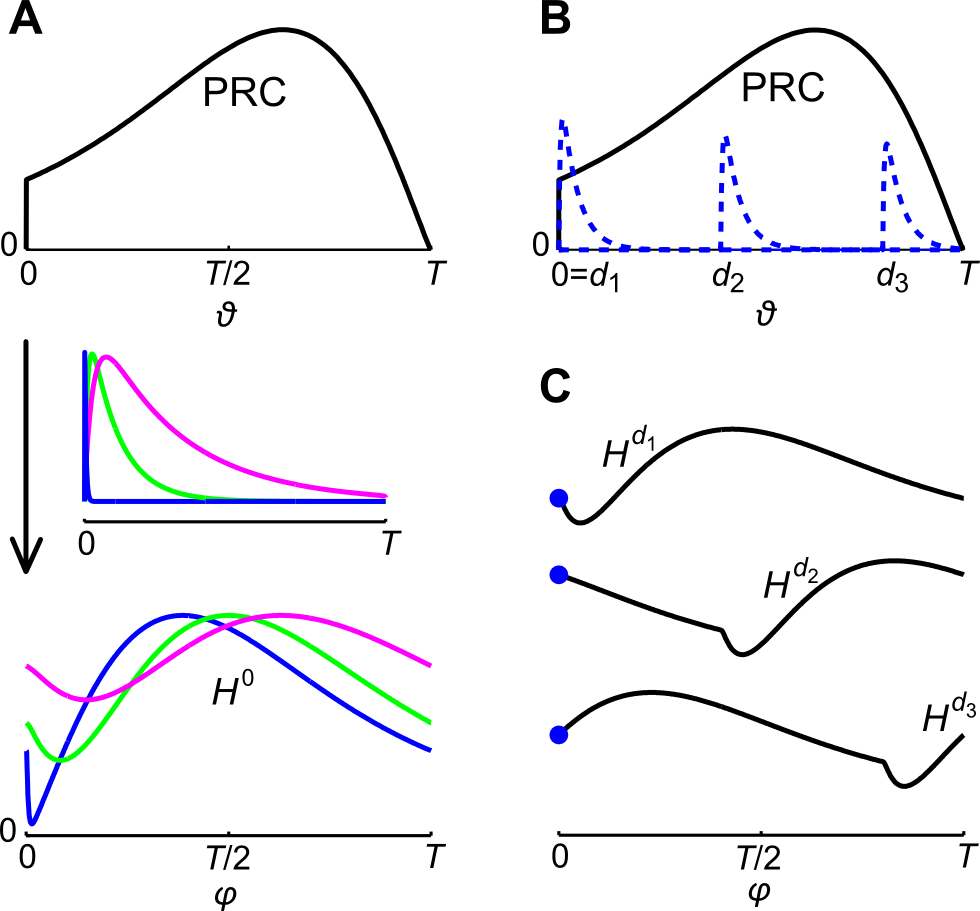}
  \end{center}  
  \caption{\textbf{Relationship between the PRC and the interaction function}. A: PRC of an aEIF neuron (top) spiking at $40~\mathrm{Hz}$ and interaction functions $H^{0}(\varphi)$ (bottom) obtained for synaptic conductances with three different sets of synaptic time constants: $\tau_r=0.01~\mathrm{ms}$, $\tau_d=0.1~\mathrm{ms}$ (blue), $\tau_r=0.25~\mathrm{ms}$, $\tau_d=2.5~\mathrm{ms}$ (green); $\tau_r=0.75~\mathrm{ms}$, $\tau_d=7.5~\mathrm{ms}$ (magenta), and $d = 0$. The synaptic current $I_{syn}$ associated with each pair of time constants (center) illustrates the three synaptic timescales relative to the period $T = 25~\mathrm{ms}$. Note that $I_{syn}$ shown here is received by the neuron at the beginning of its ISI.
B: PRC (solid black) of an aEIF neuron spiking at $40~\mathrm{Hz}$ and excitatory synaptic currents $I_{syn}$ with $\tau_r=0.1~\mathrm{ms}$, $\tau_d=1~\mathrm{ms}$ (dashed blue) received at three different phases. Assuming the input comes from a second, synchronous neuron, these phases represent three different conduction delays $d_1 = 0~\mathrm{ms}$, $d_2 = 10~\mathrm{ms}$, and $d_3 = 20~\mathrm{ms}$. Note that synaptic input received at an earlier phase causes a larger peak of $I_{syn}$, due to the smaller value $V$ of the membrane potential which leads to a larger difference $E_{syn}-V$ to the synapse's reversal potential $E_{syn}$. C: Interaction functions $H^{d_i}(\varphi)$ for pairs of neurons with the PRC shown in B, coupled by excitatory synapses with $\tau_r=0.1~\mathrm{ms}$, $\tau_d=1~\mathrm{ms}$, and delays $d_1, d_2$ and $d_3$. The values of $H^{d_i}(\varphi)$ at $\varphi = 0$ are highlighted by blue circles. The slopes of $H^{d_i}(0)$, in terms of both left and right sided limits $\mathrm{lim}_{\varepsilon \searrow 0} \, d H^{d_i}(- \varepsilon) / d \varphi$ and $\mathrm{lim}_{\varepsilon \searrow 0} \, d H^{d}(\varepsilon) / d \varphi$, indicate whether the synchronous states are stable or unstable (see main text).}
  \label{fig_4}
\end{figure}

\noindent A synaptic current with finite rise and decay times causes an additional rightwards shift and a smoothing of the interaction function. The stability of the fixed point $\hat{\varphi}$ is then determined by the slope of the PRC and its discontinuity on the interval 
$(d - \hat{\varphi}, \: d - \hat{\varphi} + \varepsilon)$, where $\varepsilon > 0$ is on the order of the synaptic timescale (see Text S1 C). 
If the PRC slope is negative on this interval and its discontinuity (if occurring in the interval) is also negative, i.e. $\text{PRC}( T^- ) > \text{PRC}( 0 )$, then $\hat{\varphi}$ is stable for excitatory coupling and unstable for inhibitory coupling.
In Fig.~\ref{fig_4}A we show the effect of the synaptic timescale, i.e. $\tau_r$ and $\tau_d$, on the interaction function for a given PRC. Fig.~\ref{fig_4}B,C illustrates how the stability of the synchronous state of a neuronal pair is given by the slope of the PRC, for three different delays. The slope of the PRC is positive at 
$\vartheta = d_1^+$, $\vartheta = d_2$ and negative at $\vartheta = d_3$ and remains positive (negative) 
until $I_{syn}$ has decayed to a small value. Therefore, synchrony is unstable for delays $d_1$, $d_2$ and stable for $d_3$, indicated by the slope of $H^d$ at $\varphi = 0$, which is negative for the first two and positive for the third delay.
\\

\noindent 
\textbf{Effects of adaptation on phase locking of coupled aEIF pairs}

\noindent First, consider pairs of identical aEIF neurons with the PRCs shown in Fig.~\ref{fig_2}A,B, symmetrically coupled through instantaneous synapses ($\tau_r \searrow 0$ and $\tau_d \searrow 0$) and without conduction delays ($d = 0$). When the coupling is excitatory, the in-phase locked state (synchrony) is unstable in case of \mbox{type I} PRCs, since they have a positive ``jump" at $\vartheta = 0$, i.e. $\text{PRC}( T^- ) < \text{PRC}(0)$. Synchrony is stable for pairs with \mbox{type II} PRCs however, as $\text{PRC}( T^- ) > \text{PRC}( 0 )$.
The anti-phase locked state on the other hand is unstable because of the positive PRC slopes at $\vartheta = T/2$. In case of inhibitory coupling, synchrony is stable for \mbox{type I} pairs and the anti-phase locked state is stable for all pairs. This means, bistability of in-phase and anti-phase locking occurs for inhibitory neurons with \mbox{type I} PRCs.
\\

\begin{figure}[!htp]
  \begin{center}
  \includegraphics[width=0.70\textwidth]{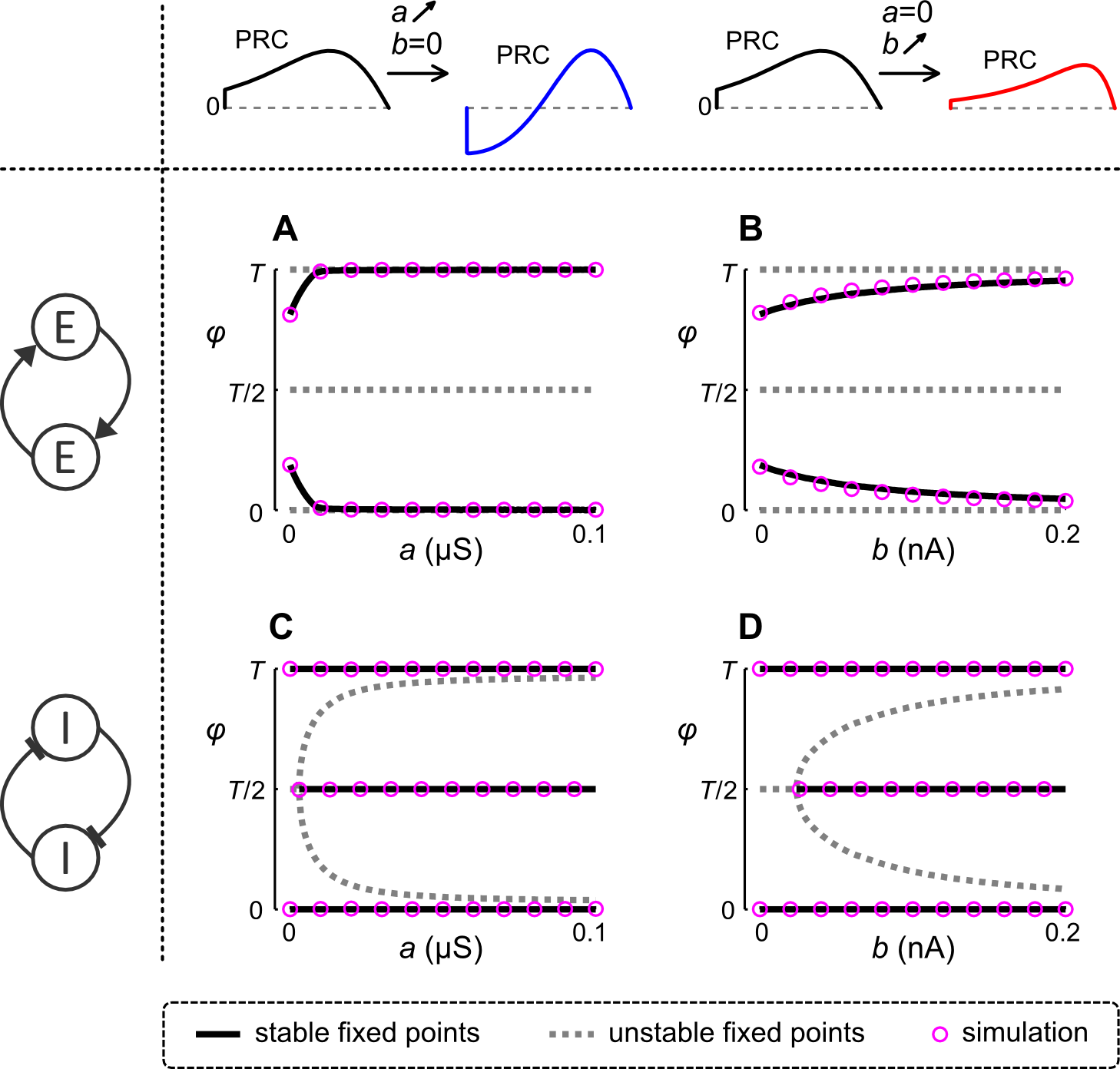}
  \end{center}  
  \caption{\textbf{Effects of adaptation on phase locked states of coupled aEIF pairs}.
    Stable (solid black) and unstable (dashed grey) phase 
    locked states of pairs of aEIF neurons spiking at $40~\mathrm{Hz}$ with identical PRCs as a function of 
    adaptation parameters. These phase locked states were 
    obtained by evaluating the interaction function. Circles denote the steady-state 
    phase differences by numerically simulating pairs of aEIF neurons according 
    to eqs.~\eqref{eq_adex_membrane_current}--\eqref{eq_adex_reset}. 
To detect bistability, the simulations were run multiple times and the pairs initialized either near in-phase or anti-phase with values of the periodic spiking trajectory.
    In A and B the neurons are coupled through excitatory, 
    in C and D through inhibitory synapses, as indicated by the diagrams on the left.
    Synaptic conductances are equal ($g_{12}=g_{21}$) and conduction delays are not 
    considered here ($d_{12}=d_{21} =: d=0$). Synaptic time constants were  $\tau_r=0.1~\mathrm{ms}$, $\tau_d=1~\mathrm{ms}$ for excitatory and  $\tau_r=0.5~\mathrm{ms}$, $\tau_d=5~\mathrm{ms}$ for inhibitory connections. In A and C, $a$ varies from 0 to 0.1~$\mathrm \mu S$ 
    with $b=0~\mathrm{nA}$, whereas in B and D, $a=0~\mathrm{\mu S}$ while $b$ varies from 
    0 to 0.2~nA. All other model parameters are given in the Methods section.
    The corresponding changes in PRCs are indicated in the top row.
  }
  \label{fig_5}
\end{figure}

\noindent Next, we consider pairs that are coupled through synaptic currents $I_{syn}$ with finite rise and decay times, as described in the Methods section. In Fig.~\ref{fig_5} we show how the stable (and unstable) phase 
locked states of pairs of neurons with symmetric excitatory (A, B) 
and inhibitory (C, D) synaptic interactions and without conduction delays change, when the 
PRCs are modified by the adaptation components $a$ and $b$. For excitatory 
pairs, stable fixed points shift towards synchrony, when $a$ or $b$ is increased. The phase differences become vanishinly small, when the PRCs switch from \mbox{type I} to \mbox{type II} due to subthreshold adaptation. Perfect synchrony is stabilized, where the PRC slopes at $\vartheta = \varepsilon$ for small $\varepsilon >0$ become negative, due to even larger values of $a$ (not shown) or lower spike frequency (see Fig.~\ref{fig_3}C--F). Neurons that have 
\mbox{type I} PRCs with a pronounced skew, as caused by spike-triggered adaptation, 
lock almost but not completely in-phase, if the adaptation is sufficiently strong. 
Inhibitory pairs on the other hand show stable synchrony independent of 
PRC type and skewness. Larger values of $a$ or $b$ lead to additional 
stabilization of the anti-phase locked state (through a subcritical pitchfork bifurcation). 
That is, strong adaptation in inhibitory pairs mediates bistability of in-phase and anti-phase locking. All phase locking predictions from the phase reduction approach are in good agreement with the results of numerically simulated coupled aEIF pairs.
\\ \\

\noindent 
\textbf{Phase locking of aEIF pairs coupled with delays}

\begin{figure}[!htp]
  \begin{center}
  \includegraphics[width=0.85\textwidth]{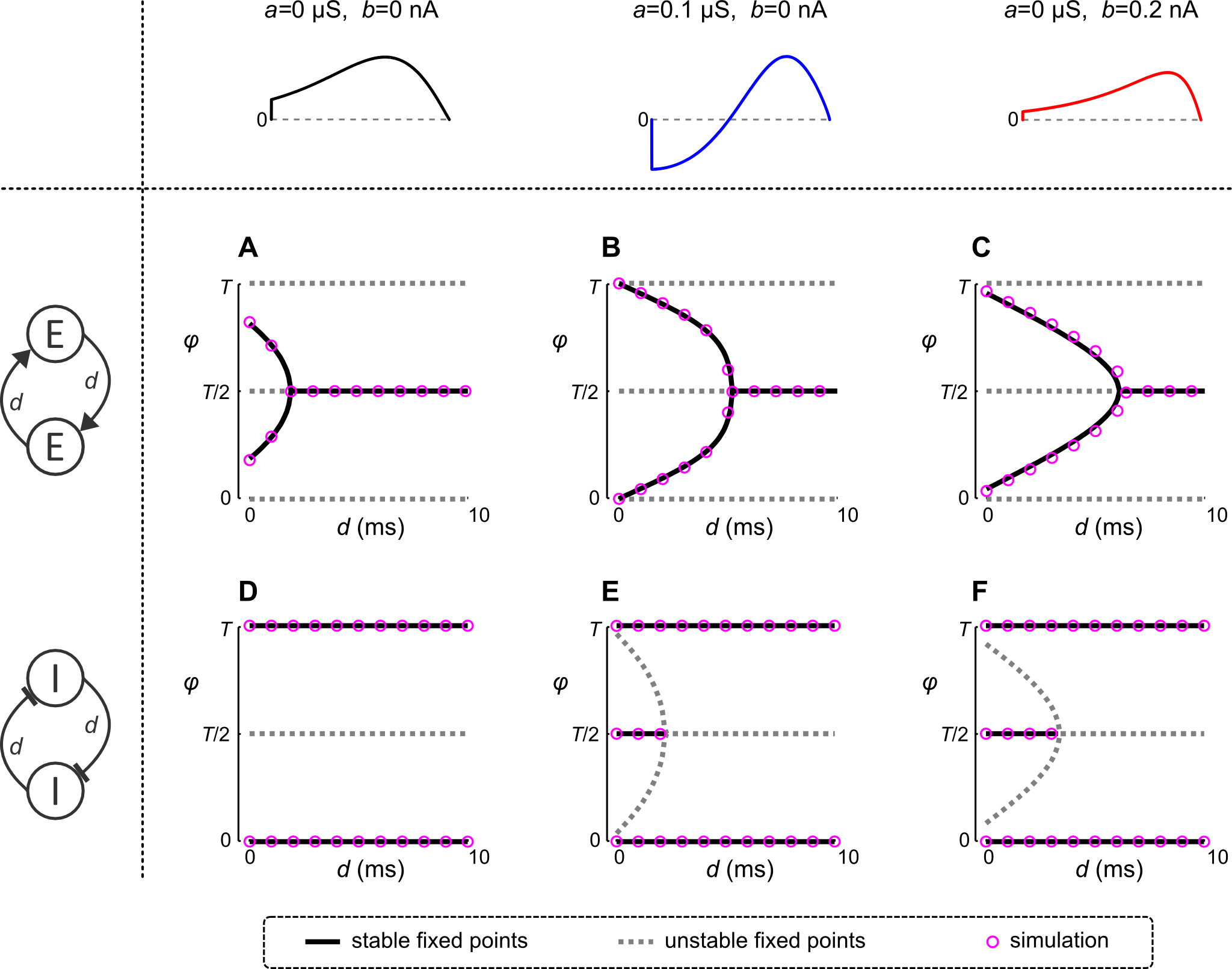}
  \end{center}  
  \caption{\textbf{Phase locking of coupled aEIF pairs with conduction delays}.
    Stable (solid black) and unstable (dashed grey) 
    phase locked states of aEIF pairs without adaptation, $a=b=0$ (A and D), and with
    adaptation, $a=0.1~\mathrm{\mu S}$, $b=0~\mathrm{nA}$ (B and E), $a=0~\mathrm{\mu S}$,
    $b=0.2~\mathrm{nA}$ (C and F), as a function of the conduction delay $d$. 
    The neurons are coupled through 
    excitatory (A-C) or inhibitory synapses (D-F) with equal conductances 
    ($g_{12}=g_{21}$). Synaptic time constants are as in 
    Fig.~\ref{fig_5}. Circles denote steady-state phase 
    differences of numerically simulated pairs of aEIF neurons. The corresponding PRCs
    are shown in the top row. $T$ was 25~ms.
  }
  \label{fig_6}
\end{figure}

\begin{figure}[!htp]
  \begin{center}
  \includegraphics[width=0.75\textwidth]{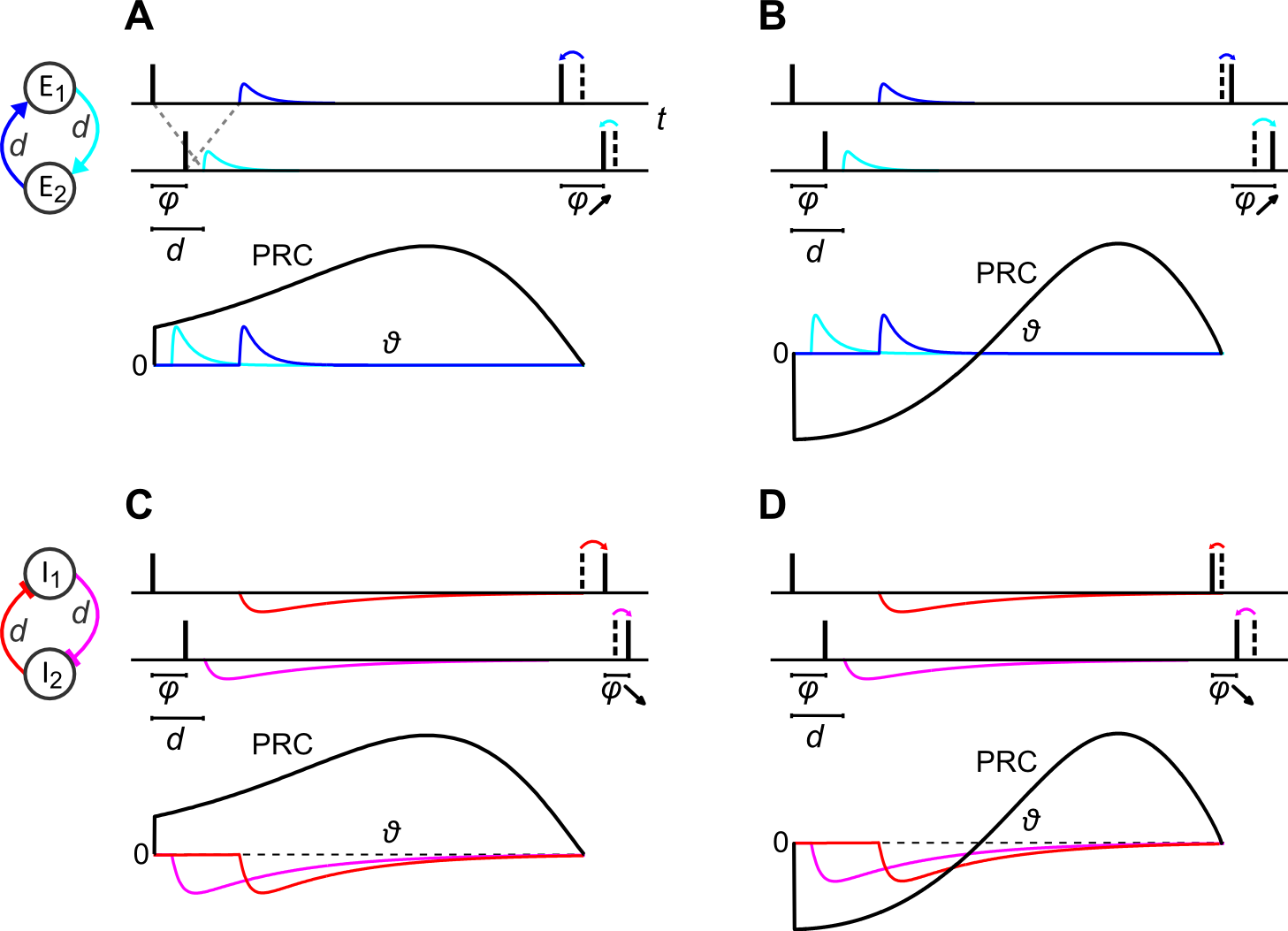}
  \end{center}  
  \caption{\textbf{Effects of conduction delays on the stability of synchrony in coupled pairs}. Spike times (solid bars) of two neurons oscillating with a small phase difference $\varphi$ and coupled through excitatory (A and B) or inhibitory synapses (C and D) with a symmetric conduction delay $d$. The PRCs of the neurons that make up each pair are displayed below. In A and C the neurons have \mbox{type I} PRCs, in B and D the PRCs are \mbox{type II}. The time (phase) at which each neuron receives a synaptic current is shown along the spike trace. 
Phase advances or delays, considering the time of input arrival and the shape of the PRC, are indicated by advanced or delayed subsequent spike times. Dashed bars indicate spike times without synaptic inputs. The consequent changes in $\varphi$ are highlighted.}
  \label{fig_7}
\end{figure}

\noindent We next investigate how phase locked states of excitatory and 
inhibitory pairs are affected by synaptic currents that involve conduction delays, considering the PRC of a 
neuron without adaptation, and two PRCs 
that represent adaptation induced by either $a$ or $b$. Neurons symmetrically coupled through 
excitatory synapses with a conduction delay do not synchronize 
irrespective of whether adaptation is present or not (Fig.~\ref{fig_6}A-C). 
Instead, stable states shift towards anti-phase locking with increasing 
mutual delays (where the anti-phase locked state eventually stabilizes by a supercritical pitchfork bifurcation).
Inhibitory pairs on the other hand synchronize for all 
conduction delays (Fig.~\ref{fig_6}D-F), but the anti-phase 
locked states of coupled inhibitory neurons with \mbox{type II} PRCs or skewed
\mbox{type I} PRCs are destabilized by the delays (by a subcritical pitchfork bifurcation).
The bistable region is larger in case of spike-triggered adaptation compared to subthreshold adaptation (Fig.~\ref{fig_6}E,F). 
Again, all stable phase locked states obtained using phase reduction are verified by numerical simulations.
Fig.~\ref{fig_7} illustrates the phenomenon that 
synchronous spiking of excitatory pairs is destabilized by the delay, 
while synchrony remains stable for inhibitory pairs. Consider two neurons 
oscillating with a small phase difference $\varphi=\vartheta_1 - \vartheta_2 > 0$ 
(neuron 1 slightly ahead of neuron 2). Then, a synaptic input 
received by neuron 2 at a delay $\varphi < d < T/2$ after neuron 1 
has spiked, arrives at an earlier phase ($\vartheta_2 = d - \varphi$) compared to the phase at which neuron 1 receives its input ($\vartheta_1 = d + \varphi$). Consequently, if the synapses are
excitatory and the PRCs \mbox{type I}, the leader neuron 1 advances its next spike by a larger amount than the
follower neuron 2 (Fig.~\ref{fig_7}A). In case of excitatory neurons and \mbox{type II} PRCs, depending on $\varphi$ and $d$, the phase of neuron 1 is advanced by a larger amount or delayed by a smaller amount than the phase of neuron 2, the latter of which is shown by the changed spike times in Fig.~\ref{fig_7}B. It is also possible that the phase of the leader neuron is advanced while that of the follower neuron is delayed. Hence, for either PRC type, $\varphi$ increases due to delayed excitatory coupling, that is, synchrony is destabilized.
For inhibitory synapses and \mbox{type I} PRCs, the leader neuron 1 delays its subsequent spike by a larger amount than the follower neuron 2 (Fig.~\ref{fig_7}C). In case of \mbox{type II} PRCs, neuron 1 experiences a weaker phase advance or stronger phase delay than neuron 1, or else the phase of neuron 1 is delayed while that of neuron 1 is advanced, depending on $\varphi$ and $d$ (Fig.~\ref{fig_7}D). Thus, delayed inhibitory coupling causes $\varphi$ to decrease towards zero for either PRC type, that is, synchrony is stabilized.
\\

\noindent \textbf{Phase locking of aEIF pairs coupled with delays and unequal synaptic strenghts}

\begin{figure}[!htp]
  \begin{center}
  \includegraphics[width=0.85\textwidth]{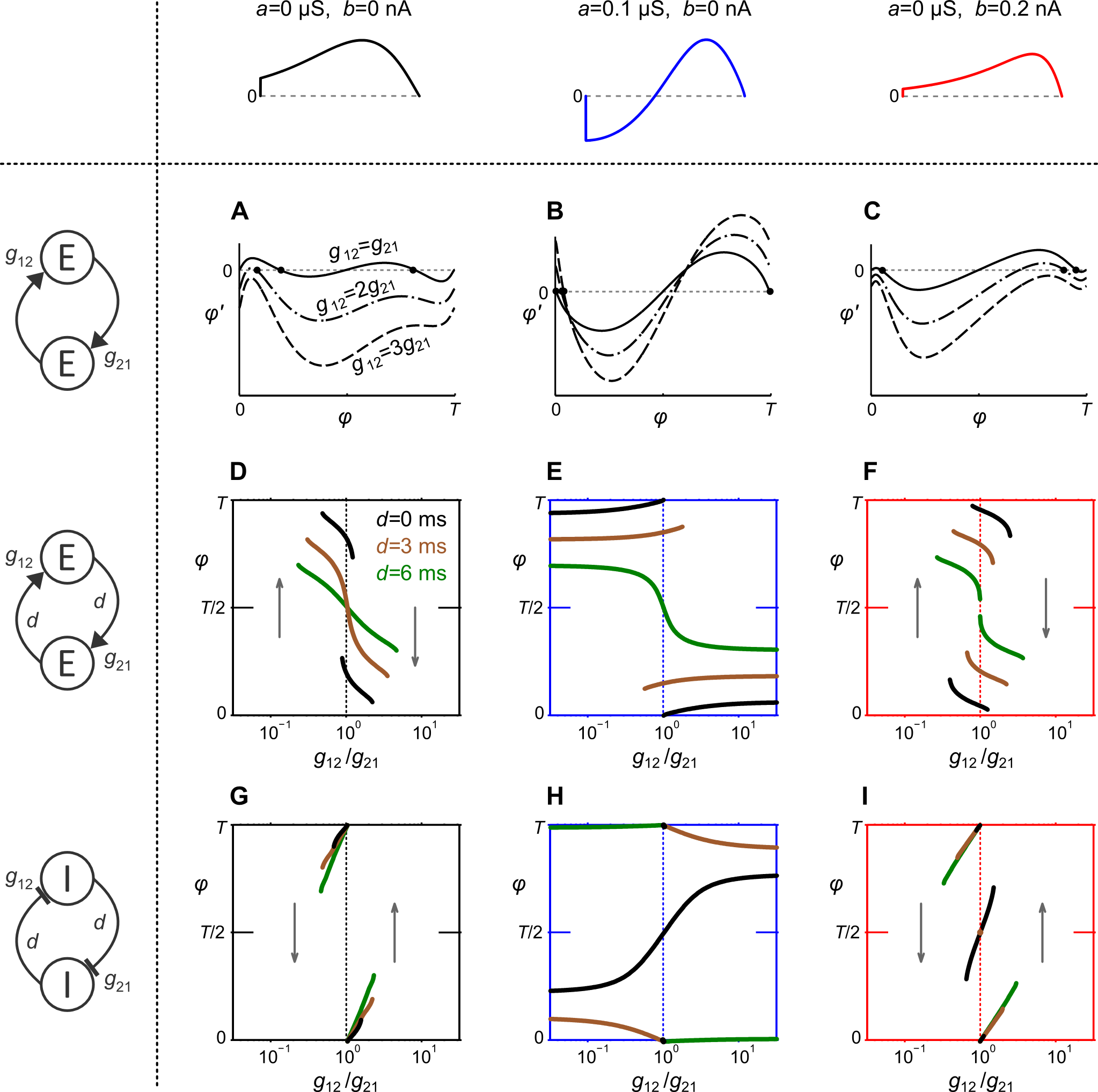}
  \end{center}  
  \caption{\textbf{Phase locking of aEIF pairs coupled with delays and 
    heterogeneous synaptic strengths}. A-C: Change of phase difference 
    $\varphi' := d \varphi / dt$ given by equation 
    \eqref{eq_scalar_phase_ode_heterogcouplings}, as a function of 
    $\varphi$ for pairs of 
    excitatory aEIF neurons coupled with different ratios of synaptic 
    conductances $g_{12}/g_{21}$ ($d=0$). Zero crossings with a 
    negative slope indicate stable phase locking and are marked by black 
    dots. Adaptation parameters of the neurons and PRCs are shown in the 
    top row. D-I: Stable phase locked states of excitatory (D-F) and 
    inhibitory (G-I) pairs as a function of the synaptic conductance ratio, 
    for three different conduction delays $d=0$, $3$ and $6~\mathrm{ms}$ (black, 
    brown, green). Unstable states are not shown for improved clarity. 
    Dashed lines denote equal synaptic strengths, grey 
    arrows indicate a continuous increase or decrease of $\varphi$ (mod $T$) 
    for ratios $g_{12}/g_{21}$ at which phase locked states do not exist (see main text).
  }
  \label{fig_8}
\end{figure}

\noindent In the following we analyze phase locking of neuronal pairs with 
unequal synaptic peak conductances $g_{12} \neq g_{21}$. Due to the linearity of the 
integral in eq.~\eqref{eq_phasenet_integral} we can substitute 
$H_{ij}^d =: g_{ij} \tilde H_{ij}^d$ in eq.~\eqref{eq_scalar_phase_ode}, 
which yields
\begin{equation} \label{eq_scalar_phase_ode_heterogcouplings}
 \frac{d \varphi}{dt} = g_{21} \tilde H_{21}^d(-\varphi) - 
      g_{12} \tilde H_{12}^d (\varphi).
\end{equation}
By setting eq.~\eqref{eq_scalar_phase_ode_heterogcouplings} to zero, we obtain 
the condition eq.~\eqref{eq_phaselockedstates_heterogcoupling} for the existence 
of phase locked states,
\begin{equation} \label{eq_phaselockedstates_heterogcoupling}
  \frac{g_{12}}{g_{21}} = \frac{\tilde H_{21}^d(-\varphi)}{\tilde H_{12}^d (\varphi)}.
\end{equation}
Phase locked states therefore only exist if the ratio of conductances 
$g_{12}/g_{21}$ is not larger than the maximum of the 
periodic function $R(\varphi) := 
  \tilde H_{21}^d(-\varphi) / \tilde H_{12}^d (\varphi)$. 
This upper bound primarily depends on the type of the PRCs and the
synaptic time constants. In case of \mbox{type I} PRCs, 
$\max_{\varphi} R(\varphi)$ is limited because  the minimum of 
$| \tilde H_{ij}^d(\varphi) |$ is positive. $\tilde H_{ij}^d(\varphi)$ is either
positive (for excitatory synapses) or negative (for inhibitory synapses) for all $\varphi$. $\max_{\varphi} R(\varphi)$ is small for slow synapses, since the slower the synaptic rise and decay times, the larger $\min_{\varphi} | \tilde H_{ij}^d(\varphi) |$, see Fig.~\ref{fig_4}A. For a \mbox{type II} PRC on the other 
hand, this minimum is zero (unless the negative lobe of the PRC is small and the 
synapse slow), from which follows that
$\max_{\varphi} R(\varphi) \to \infty$. The effects of heterogeneous synaptic 
strengths on phase locking of neuronal pairs without adaptation, as well as
either adaptation parameter increased, are shown in Fig.~\ref{fig_8}. For 
excitatory pairs coupled without a conduction delay it is illustrated, how the right hand 
side of eq.~\eqref{eq_scalar_phase_ode_heterogcouplings} changes when the 
coupling strengths are varied (A-C). In addition, stable phase locked states 
of excitatory and inhibitory pairs coupled through synapses with various mutual
conduction delays ($d=0$, $3$, or $6~\mathrm{ms}$) are displayed as a function of
$g_{12}/g_{21}$ (D-I). When the ratio of conductances $g_{12}/g_{21}$ 
is increased, the zero crossings of $d \varphi / dt$ given by eq.~\eqref{eq_scalar_phase_ode_heterogcouplings}, i.e. phase locked states, disappear 
for neurons with \mbox{type I} PRCs (through a SN bifurcation). $\varphi$ then continuously increases (or decreases) (mod $T$) as shown by the dashed curves (without roots) in Fig.~\ref{fig_8}A,C and indicated by the arrows in Fig.~\ref{fig_8}D,F,G,I. This means, the spike frequency of one neuron becomes faster than that of the other neuron. Neurons with \mbox{type II} PRCs on the other hand have 
stable phase locked states even for diverging coupling strengths. Bistability of two phase locked states can occur for a ratio $g_{12}/g_{21}$ close to one (equal coupling strengths), depending on the PRC and the delay.
Synchronization of excitatory-inhibitory pairs is not considered in this paper.
It should be noted however, that if both neurons have \mbox{type I} PRCs, phase locking is not possible, 
irrespective of the ratio of coupling strengths. In this case, one interaction function is
strictly positive and the other strictly negative and thus, the condition 
\eqref{eq_phaselockedstates_heterogcoupling} for fixed points of eq.~\eqref{eq_scalar_phase_ode_heterogcouplings} cannot be fulfilled.

\subsection*{Synchronization and clustering in aEIF networks}

\begin{figure}[!htp]
  \begin{center}
  \includegraphics[width=1.0\textwidth]{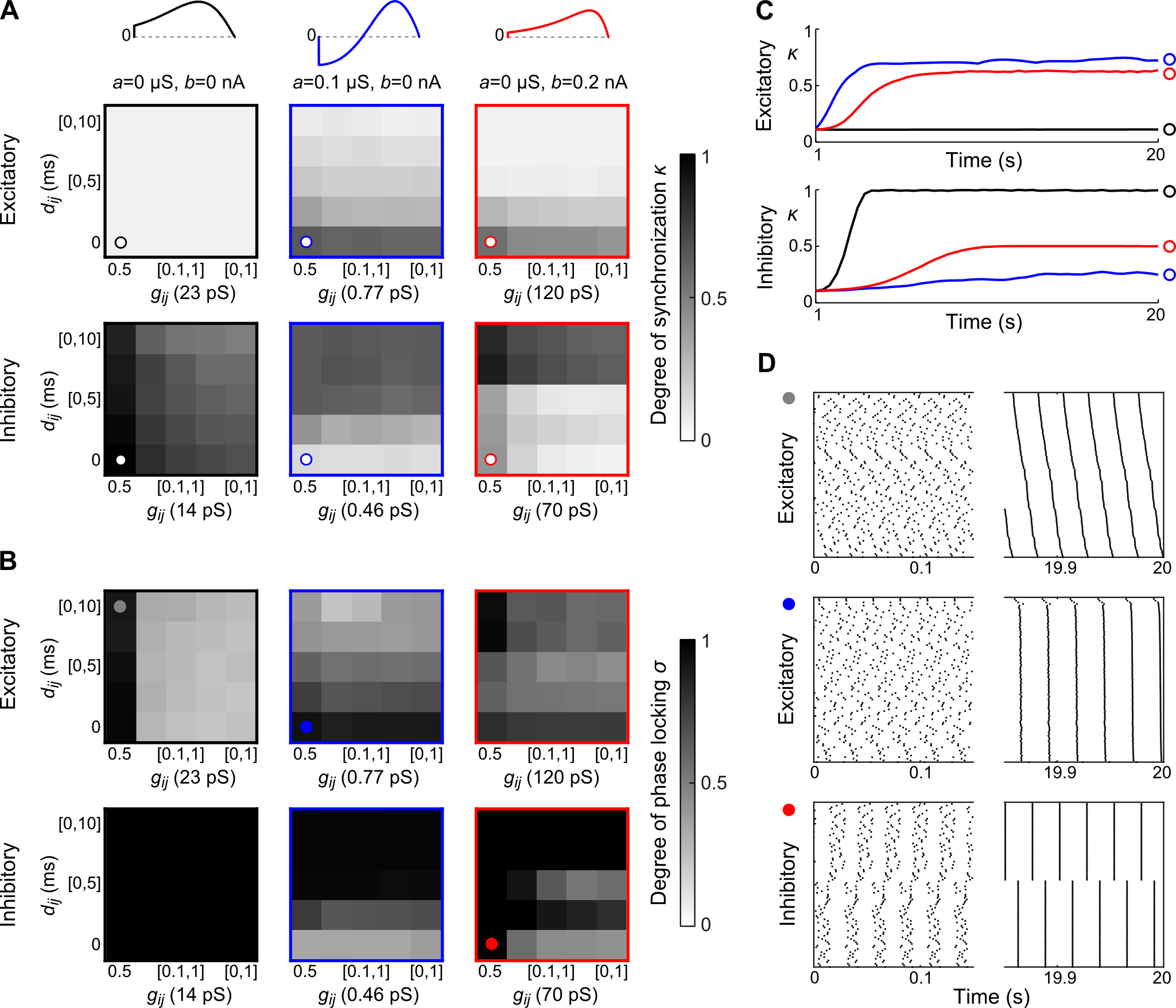}
  \end{center}  
  \caption{\textbf{Impact of adaptation on the behavior of aEIF networks}. 
    Degree of network synchronization $\kappa$ (A) and phase locking $\sigma$ (B)
    of $N=100$ aEIF neurons without adaptation, 
    $a=b=0$ (black frame) and either adaptation component, respectively, 
    $a=0.1~\mathrm{\mu S}$, $b=0~\mathrm{nA}$ (blue frame), $a=0~\mathrm{\mu S}$, 
    $b=0.2~\mathrm{nA}$ (red frame), driven to 40~Hz spiking, all-to-all coupled without 
    self-feedback, for various 
    conduction delays and synaptic conductances. $d_{ij}$ and $g_{ij}$ are 
    random (uniformly distributed) in the indicated intervals. Specifically, 
    $d_{ij}=0$, $d_{ij} \in [0,2.5],\, [0,5],\, [0,7.5],\, [0,10]$ and 
    $g_{ij} = 0.5$, $g_{ij} \in [0.2,1], \,[0.1,1], \,[0.02,1], \,[0,1]$,
    with units in parenthesis. 
    The PRCs of the three neuron types described above are shown in the top row. 
    C: Time course of $\kappa$ for networks without delays and equal synaptic strengths, 
    as indicated by the symbols in A. Each $\kappa$ and $\sigma$ 
    value represents an average over three simulation runs. 
    D: Raster plots for neuron and network 
    parameters as indicated by the symbols in B, where the neurons in the columns are
    sorted according to their last spike time.
  }
  \label{fig_9}
\end{figure}

In order to examine how the behavior of pairs of coupled phase neurons 
relates to networks of spiking neurons, we performed numerical simulations 
of networks of oscillating aEIF neurons without adaptation and with either a
subthreshold or a spike-triggered adaptation current, respectively, and 
analyzed the network activity. The neurons were all either excitatory or inhibitory 
and weakly coupled. Fig.~\ref{fig_9} 
shows the degree of synchronization $\kappa$ (A, C) and the degree of phase locking $\sigma$ (B) for these networks considering equal as well as heterogeneous 
conduction delays and synaptic conductances. An increase of either adaptation parameter ($a$ or $b$) leads to increased $\kappa$ in networks of excitatory neurons with short delays. It can be recognized however, that $\kappa$ increases to larger values and this high degree of synchrony seems to be more robust against heterogeneous synaptic strengths, when the neurons are equipped with a subthreshold adaptation current (Fig.~\ref{fig_9}A,C). 
These effects correspond well to those of the adaptation components $a$ and $b$ on 
synchronization of pairs, presented in the previous section. Parameter regimes (w.r.t. $a,b,d_{ij}$ and $g_{ij}$) that cause stable in-phase or near in-phase locking of pairs, such as subthreshold adaptation in case of short delays or spike-triggered adaptation for short delays and coupling strength ratios close to one 
(Fig.~\ref{fig_6}A-C and Fig.~\ref{fig_8}D-F), lead to synchronization, indicated by large $\kappa$ values, in the respective networks. 
Networks of non-adapting 
excitatory neurons remain asynchronous as shown by the low $\kappa$ values. 
For equal synaptic strengths, these networks settle into splay states 
where the neurons are pairwise phase locked, with uniformly distributed phases 
(Fig.~\ref{fig_9}B,D). When the delays are large enough and the synaptic strengths equal, splay states also occur in networks of neurons with large $b$, indicated by low $\kappa$ and high $\sigma$ values in Fig.~\ref{fig_9}A,B. 
As far as inhibitory networks are concerned, non-adapting neurons synchronize, without delays or with random delays of up to 10~ms. Furthermore, synchrony in these networks is largely robust against heterogeneities in the coupling strengths (Fig.~\ref{fig_9}A). Networks of inhibitory neurons with subthreshold adaptation only show synchronization and pairwise locking for larger delays (i.e. $d_{ij}$ random in $[0,5~\mathrm{ms}]$ or larger). Spike-triggered adaptation promotes clustering of the network into two clusters, where the neurons within a cluster are in synchrony, as long as the delays are small. These cluster states seem to be most robust against heterogeneous synaptic strengths when the delays are small but not zero.
For larger delays, inhibitory neurons of all three types (with or without adaptation) synchronize, in a robust way against unequal synaptic strengths. 
The behaviors of inhibitory networks are consistent with the 
phase locked states found in pairs of inhibitory neurons (Fig.~\ref{fig_6}D-F).
Particularly, stable synchronization of pairs with larger conduction delays and the 
bistability of in-phase and anti-phase locking of pairs with spike-triggered adaptation for smaller delays, nicely 
carry over to networks. In the former case, synchrony of pairs relates to network synchrony, in the latter case, bistability of in-phase and anti-phase locking of individual pairs can explain the observed two cluster states. Note that bistability of in-phase and anti-phase locking is also shown for inhibitory pairs with subthreshold adaptation and $d=0~\mathrm{ms}$. In this case however, the slope of $H_{\Delta}(\varphi)$ at $\varphi = T/2$ is almost zero (not shown), which might explain why the corresponding networks do not develop two-cluster states. The behavior of all simulated networks 
does not critically depend on the number of neurons in the network, as we obtain qualitatively similar results for network sizes changed to $N=50$ and $N=200$ (not shown).
The numerical simulations demonstrate that stable phase
locked states of neural pairs can be used to predict the behavior of larger networks.

\subsection*{Synchronization properties of Traub neurons with adaptation currents $I_m$, $I_{ahp}$}

\noindent To understand the biophysical relevance of the subthreshold and spike-triggered
adaptation parameters, $a$ and $b$, in the aEIF model, we compare them with the adaptation currents $I_m$ and $I_{ahp}$ in a variant of the Hodgkin-Huxley type Traub model neuron. Specifically, in this section we investigate the effects of the low- and high-threshold currents $I_m$ and $I_{ahp}$, respectively, on spiking 
behavior, \mbox{$F$-$I$} curves and PRCs of single neurons, and on synchronization of pairs and networks, using the Traub model, and compare the results with those of the previous two sections. It should be stressed, that the aEIF model was not fit to the Traub model in this study. Therefore, the comparison of how adaptation currents affect SFA, PRCs and synchronization in both models, are rather qualitative than quantitative.
\\

\noindent \textbf{PRC characteristics of Traub neurons}

\begin{figure}[!htp]
  \begin{center}
  \includegraphics[width=0.9\textwidth]{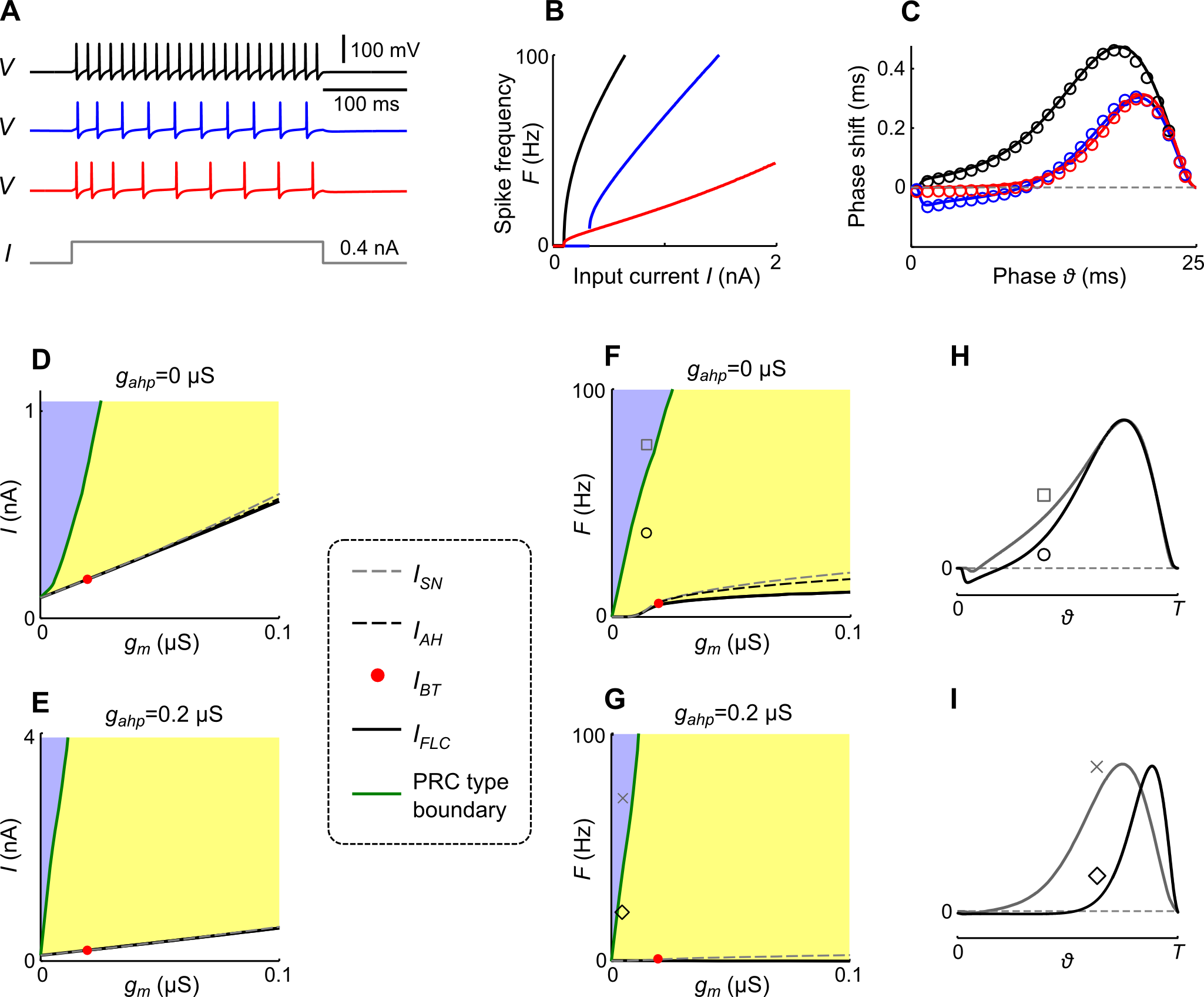}
  \end{center} 
  \caption{\textbf{Effects of adaptation on spiking dynamics, 
    $\mathbf F$-$\mathbf I$ 
    curves, PRCs and bifurcation currents of Traub model neurons}. 
    A: Membrane potential $V$ of Traub model neurons without adaptation, 
    $g_m=g_{ahp}=0~\mathrm{\mu S}$ (black), $I_m$-mediated, 
    $g_m = 0.1~\mathrm{\mu S}$ (blue) and $I_{ahp}$-mediated adaptation, 
    $g_{ahp} = 0.1~\mathrm{\mu S}$ (red), in response to step currents 
    $I$, 
    B: the corresponding \mbox{$F$-$I$} curves, and 
    C: the corresponding PRCs.
    Solid lines in C denote the PRCs, 
    calculated with the adjoint method and scaled by 0.2~mV. Open circles denote the results of numerical simulations of eqs.~\eqref{eq_traub_currentbalance}--\eqref{eq_traub_caconcentration} with 0.2~mV
    perturbations at various phases. 
    D,E: Rheobase current $I_{FLC}$ (solid black), $I_{SN}$ (dashed grey) and 
    $I_{AH}$ (dashed black), as a function of $g_m$, for $g_{ahp}=0~\mathrm{\mu S}$ 
    (left) and $g_{ahp}=0.2~\mathrm{\mu S}$ (right). $I_{SN}$ and $I_{AH}$ converge at
    $I_{BT}$ marked by the red dot. The input current indicated 
    by the green curve separates \mbox{type I} and \mbox{type II} PRC regions (blue and 
    yellow, respectively). F,G: Spike frequencies $F$ according to the input
    currents $I$ in D and E. H,I: PRCs for parametrizations as indicated in F and G
    (with $I$ corresponding to $F$), scaled to the same period $T$.
    All other model parameters are provided in the Methods section.
  }
  \label{fig_10}
\end{figure}

\noindent Without adaptation, $g_m = g_{ahp} = 0$ (hence $I_m = I_{ahp} = 0$), the model exhibits tonic spiking in response to a rectangular current 
pulse (Fig.~\ref{fig_10}A). When either adaptation current is present, that is the conductance $g_m$ or $g_{ahp}$ is increased to $0.1~\mathrm{\mu S}$, the membrane voltage trace reveals SFA. Note that $I_{ahp}$ causes stronger differences in subsequent ISIs after stimulus onset, when comparing the $V$-traces of neurons with either adaptation conductance set to $0.1~\mathrm{\mu S}$. The \mbox{$F$-$I$} curves in Fig.~\ref{fig_10}B indicate
that the presence of $I_m$ predominantly has a subtractive effect on the neuron's \mbox{$F$-$I$} curve and 
gives rise to \mbox{class II} excitability. The presence of $I_{ahp}$ on the other hand flattens the \mbox{$F$-$I$} curve, 
in other words its effect is divisive. Furthermore, an increase of $I_m$ changes a \mbox{type I} PRC to 
\mbox{type II}, whereas increased $I_{ahp}$ reduces its amplitude at early phases and skews its 
peak to the right (Fig.~\ref{fig_10}C). Evidently, the effects of $I_m$ and 
$I_{ahp}$ on SFA, \mbox{$F$-$I$} curves and PRCs of Traub neurons are consistent with the effects of the adaptation parameters $a$ and $b$ in aEIF neurons (Figs.~\ref{fig_1},~\ref{fig_2}).
\\

\noindent We further show how the PRC characteristics caused by the 
adaptation currents depend on the injected current $I$, hence the spike frequency $F$, and the bifurcation type of the rest-spiking transition (Fig.~\ref{fig_10}D-I).  
An increase of $I$ reduces the effects of $I_{m}$ and 
$I_{ahp}$ on the PRC. That means, at higher frequencies $F$, larger levels of $I_m$ and $I_{ahp}$ 
are required to obtain \mbox{type II} and skewed PRCs, respectively. This frequency 
dependence of adaptation current-mediated changes of the PRC is similar in both neuron models (Figs.~\ref{fig_3},~\ref{fig_10}D-I). Note, that in the Traub model a 
rather low value of $g_m$ ($25~\mathrm{nS}$) is sufficient to guarantee a \mbox{type II} PRC for spike frequencies of up to $100~\mathrm{Hz}$ (Fig.~\ref{fig_10}F,G), compared to the 
aEIF model, where a much larger value of $a$ ($> 0.1~\mathrm{\mu S}$) would be 
necessary (Fig.~\ref{fig_3}C,D). As far as the bifurcation structures of both models are concerned, an increase of the low-threshold adaptation parameters $g_m$ and $a$ has a 
comparable effect in the Traub and the aEIF models, respectively,
changing the transition from rest to spiking from a SN via a BT to an AH 
bifurcation. The exact conductance values at which this change, i.e. the BT bifurcation, occurs, differ ($g_m = 0.02~\mathrm{\mu S}$ for the Traub model and 
$a = 0.001~\mathrm{\mu S}$ for the aEIF model).
\newline 

\noindent \textbf{Synchronization of coupled Traub neurons}

\begin{figure}[!htp]
  \begin{center}
  \includegraphics[width=0.65\textwidth]{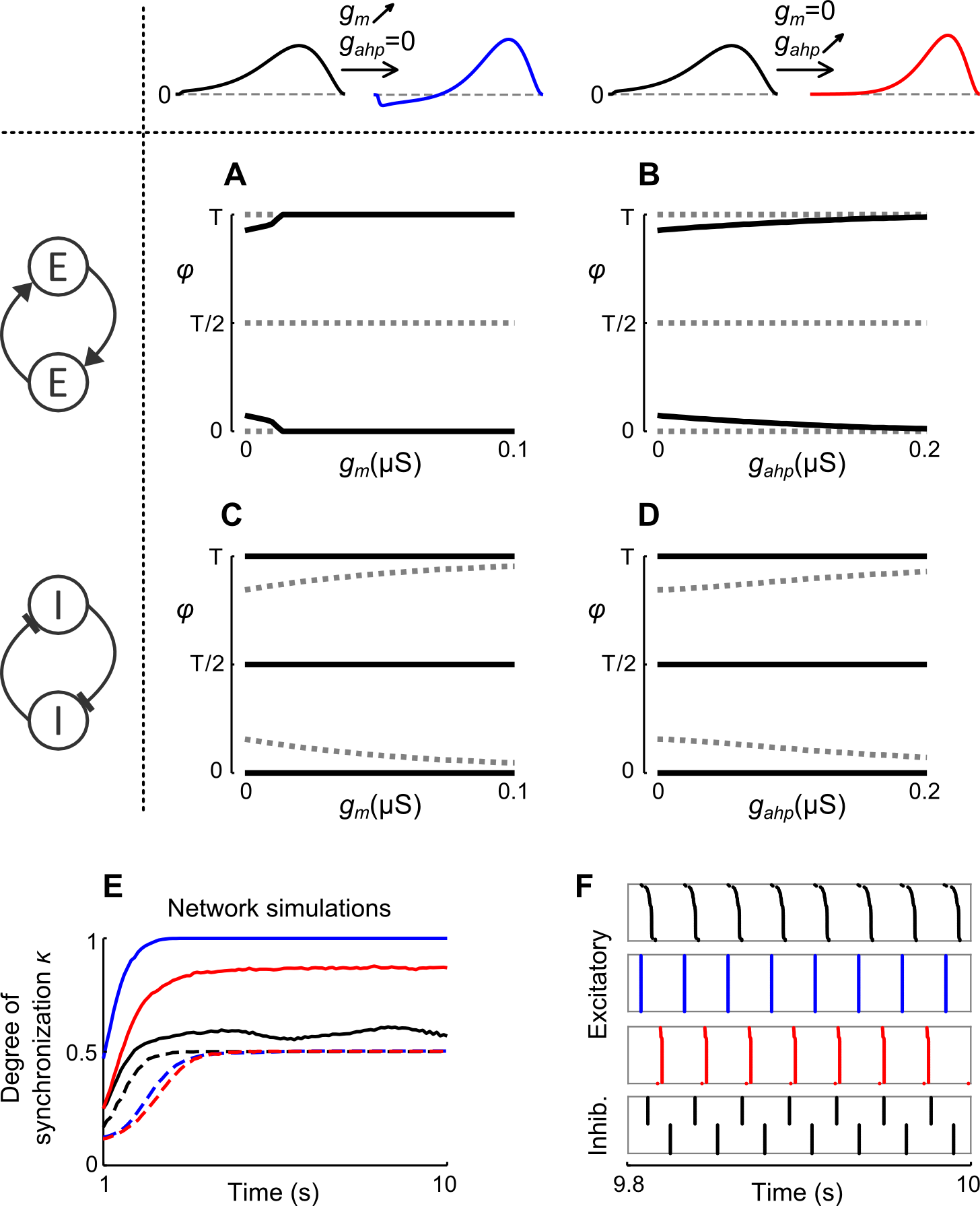}
  \end{center}  
  \caption{\textbf{Influence of adaptation on synchronization properties of 
    Traub model neurons}. A-D: Stable (solid black) and unstable (dashed grey) 
    phase locked states of coupled pairs of Traub neurons with identical PRCs, 
    as a function of conductances $g_m$ and $g_{ahp}$, respectively.
    Corresponding changes in PRCs are displayed in the top row.
    The neurons are coupled through excitatory or inhibitory synapses
    as indicated
    by the diagrams on the left, with equal synaptic strengths, $g_{12}=g_{21}$ and 
    $d=0$. E: Network synchronization $\kappa$ over time, of $N=50$ coupled 
    excitatory (solid) and inhibitory (dashed) Traub neurons 
    without, $g_m=g_{ahp}=0~\mathrm{\mu S}$ (black) or with adaptation, 
    $g_m=0.1~\mathrm{\mu S}$, $g_{ahp}=0~\mathrm{\mu S}$ (blue) and 
    $g_m=0~\mathrm{\mu S}$, $g_{ahp}=0.2~\mathrm{\mu S}$ (red), driven to 40~Hz
    spiking.
    The neurons are all-to-all coupled with equal 
    synaptic conductances, $g_{ij}=0.06~\mathrm{nS}$ (black and blue), $g_{ij}=0.18~\mathrm{nS}$ (red),
    but without self-feedback, $g_{ii}=0$, and conduction delays, $d_{ij}=0$.
    F: Raster plots showing the spike times during the last 
    200~ms for the three excitatory networks and the network of inhibitory neurons
    without adaptation (bottom). The neurons in the columns are sorted according to
    their last spike time.
  }
  \label{fig_11}
\end{figure}

\noindent  We show the effects of the adaptation currents $I_m$ and $I_{ahp}$ on phase locked states 
of pairs of Traub neurons symmetrically coupled without conduction delays in Fig.~\ref{fig_11}A-D. Excitatory pairs of neurons without adaptation 
phase lock 
with a small phase difference. Low levels of $I_m$ are sufficient to stabilize in-phase locking, by turning the PRC from \mbox{type I} to II (Fig.~\ref{fig_11}A), while an increase of $I_{ahp}$ reduces the locked phase difference to almost but not exactly zero, that is, near in-phase locking, by skewing the PRC (Fig.~\ref{fig_11}B). 
Inhibitory synaptic coupling produces bistability of in-phase (synchrony) and anti-phase locking (anti-synchrony) for pairs of neurons without adaptation or either adaptation current increased (Fig.~\ref{fig_11}C,D). Note that the domain of attraction of the anti-synchronous state 
grows with increasing $I_m$ or $I_{ahp}$, while that of the synchronous state shrinks.
In contrast to the aEIF model, this bistability also occurs for neurons without an adaptation current (compare Figs.~\ref{fig_5}C,D,~\ref{fig_11}C,D).
\\
 
\noindent The effects of $I_m$ or $I_{ahp}$ on synchronization of networks of Traub neurons coupled without conduction delays and equal synaptic strengths, are shown in Fig.~\ref{fig_11}E,F. In correspondence with the effects on pairs, 
$I_m$ and $I_{ahp}$ promote synchronization of excitatory networks, shown by the course of network synchronization measure $\kappa$ over time (Fig.~\ref{fig_11}E). The mean values of phase locking measure $\sigma$ are 0.26 for nonadapting neurons and 0.98 for networks where either adaptation current is increased. An increased adaptation current $I_m$ leads to larger $\kappa$ values, compared to an increase of $I_{ahp}$, which is similar to the aEIF networks where increased $a$ causes larger $\kappa$ values than an increase of $b$ (compare Figs.~\ref{fig_9}C,~\ref{fig_11}E). In contrast to networks of excitatory aEIF neurons without adaptation, which develop splay states, $\kappa$ values of nonadapting excitatory Traub neuron networks increase to about 0.5, while low $\sigma$ values indicate poor phase locking, hence splay states do not occur (Fig.~\ref{fig_11}F).
Networks of inhibitory neurons organize into clusters, indicated by $\kappa$ values that converge to 0.5 (Fig.~\ref{fig_11}E) and large $\sigma$ values (0.96 without adaptation, 0.94 for either $I_m$ or $I_{ahp}$ increased). Particularly, clustering into two clusters was revealed by the raster plots, see Fig.~\ref{fig_11}F. These two-cluster states of networks can be explained by the bistability of synchrony and anti-synchrony of individual pairs. Clustering emerges for all three types of Traub neurons, with and without adaptation, as opposed to networks of inhibitory aEIF neurons, where cluster states only occur in case of spike-triggered adaptation (Fig.~\ref{fig_9}).
Considering the collective behavior of coupled excitatory neurons, the synchronizing 
effects of $I_m$ and $I_{ahp}$ in the Traub model are comparable to those of the adaptation components $a$ and $b$ in the aEIF model.

\section*{Discussion}

In this work we studied the role of adaptation in the aEIF model as an endogenous neuronal mechanism that controls network dynamics. We described the effects of subthreshold and spike-triggered adaptation currents on the PRC in dependence of spike frequency. To provide insight into the synchronization tendencies of coupled neurons, we applied a common phase reduction technique and used the PRC to describe neuronal interaction \cite{Kuramoto1984, Ermentrout2010}. For pairs of coupled oscillating neurons we analyzed synchrony and phase locking under consideration of conduction delays and heterogeneous synaptic strengths. We then performed numerical simulations of aEIF networks to examine whether the predicted behavior of coupled pairs relates to the activity of larger networks. 
Finally, to express the biophysical relevance of the elementary subthreshold and spike-triggered adaptation mechanisms in the aEIF model, we compared their effects with those of the adaptation currents $I_m$ and $I_{ahp}$ in the high-dimensional Traub neuron model, on single neuron as well as network behavior.
\\

\noindent Conductance $a$, which mostly determines the amount of adaptation current in absence of spikes, that is, subthreshold, qualitatively changes the rest-spiking transition of an aEIF neuron, from a SN to an AH via a BT bifurcation as $a$ increases. Thereby the neuron's excitability, as defined by the $F$-$I$ curve, and its PRC, are turned from \mbox{class I} to \mbox{class II}, and \mbox{type I} to \mbox{type II}, respectively. A similar effect of a slow outward current that acts in the subthreshold regime on the PRC has recently been shown for a two-dimensional quadratic non-leaky integrate-and-fire (QIF) model derived from a normal form of a dynamical model that undergoes a BT bifurcation \cite{Ermentrout2010, Ermentrout2012}. The relation between the PRC and the bifurcation types has further been emphasized by Brown et al. \cite{Brown2004} who analytically determined PRCs for bifurcation normal forms and found \mbox{type I} and II PRC characteristics for the SN and AH bifurcations, respectively.
A spike-triggered increment $b$ of adaptation current does not affect the bifurcation structure of the aEIF model and leaves the excitability class unchanged. When $a$ is small such that the model is in the SN bifurcation regime, an increase of $b$ cannot change the PRC type. In the AH bifurcation regime, $b$ substantially affects the range of input current for which the PRC is \mbox{type II} but causes only a small change in the corresponding frequency range. Furthermore, spike-triggered adaptation strongly influences the skew of the PRC, shifting its peak towards the end of the ISI for larger values of $b$. Such a right-skewed PRC implies that the neuron is most sensitive to synaptic inputs that are received just before it spikes. Similar effects of spike-triggered negative feedback with slow decay on the skew of the PRC have been reported for an extended QIF model \cite{Ermentrout2001, Gutkin2005, Ermentrout2010, Ermentrout2012}.  
\\

\noindent PRCs determine synchronization properties of coupled oscillating neurons. 
When the synapses are fast compared to the oscillation period, the stability of the in-phase and anti-phase locked states (which always exist for pairs of identical neurons) can be ``read off'' the PRC for any mutual conduction delay, as we have demonstrated.
A similar stability criterion that depends on the slopes of the PRCs at the phases at which the inputs are received has recently been derived for pairs of pulse-coupled oscillators \cite{Woodman2011}. Under the assumption of pulsatile coupling, the effect of a synaptic input is required to dissipate before the next input is received. In principle, the synaptic current can be strong, but it must be brief such that the perturbed trajectory returns to the limit cycle before the next perturbation occurs \cite{Smeal2010}. 
\\

\noindent We have shown that, as long as synaptic delays are negligible and synaptic strengths equal, excitatory pairs synchronize if their PRCs are \mbox{type II}, as caused by $a$, and lock almost in-phase if their PRCs are \mbox{type I} with a strong skew, as mediated by $b$. Inhibitory pairs synchronize in presence of conduction delays and show bistability of in-phase and anti-phase locking for small delays, particularly in case of skewed PRCs. 
Conduction delays and synaptic time constants can affect the stability of synchrony in a similar way, by producing a lateral shift of the interaction function $H^d(\varphi)$, as shown in Fig.~\ref{fig_4}. Note however, that the synaptic timescale has an additional effect on the shape of $H^d(\varphi)$, smoothing it for slower synaptic rise and decay times.  
We have further demonstrated that heterogeneity in synaptic strengths desynchronizes excitatory and inhibitory pairs and leads to phase locking with a small phase difference in case of \mbox{type II} PRCs and small delays. While neurons with \mbox{type II} PRCs have stable phase locked states even for large differences in synaptic strengths, pairs of coupled neurons with \mbox{type I} PRCs are only guaranteed to phase lock when the synaptic strengths are equal. Similar effects of heterogeneous synaptic conductances have recently been observed in a computational study of weakly coupled Wang-Buszaki and Hodgkin-Huxley neurons (with \mbox{class I} and II excitability, respectively) \cite{Bradley2011}.
It should be noted that the synaptic (rise and decay) time constants alter the shape of the interaction function, while the conduction delay produces a lateral shift of the function. Both, synaptic timescale as well as delays however, affect the stability of the synchronous state in a similar way.
\\

\noindent The activity of larger aEIF networks, simulated numerically, is consistent with the predictions of the behavior of pairs. In fact, knowledge on phase locking of coupled pairs helps to explain the observed network states. Both adaptation mediated PRC characteristics, i.e. a negative lobe or a pronounced right skew, favor synchronization in networks of excitatory neurons, in agreement with previous findings 
\cite{Hansel1995, Crook1998, Ermentrout2001}. This phenomenon only occurs when the conduction delays are negligible. 
It has been shown previously that synchrony in networks of excitatory oscillators becomes unstable when considering coupling with delays \cite{Ernst1995, Ermentrout2009}. We have demonstrated that increased conduction delays promote asynchrony in excitatory networks, with or without adaptation currents. 
Inhibitory neurons on the other hand are able to synchronize spiking in larger networks for a range of conduction delays. This provides support to the hypothesis that inhibitory networks play an essential role in generating coherent brain rhythms, as has been proposed earlier 
\cite{Wang1996, Bartos2007}, \cite{Wang2010} for review. Inhibition rather than excitation has been found to generate neuronal synchrony particularly in case of slow synaptic rise and decay \cite{VanVreeswijk1994, Hansel1995, Jeong2007}, and in the presence of conduction delays as has recently been shown experimentally \cite{Wang2012}. 
In regimes that lead to 
bistability of in-phase and anti-phase locking according to our analysis of pairs, the simulated networks break up into two clusters of synchronized neurons.
Recently it has been shown that a stable two cluster state of pulse coupled neural oscillators can exist even when synchrony of individual pairs is unstable \cite{Chandrasekaran2011}.
Such cluster states have been invoked to explain population rhythms measured in vitro, where the involved neurons spike at about half of the population frequency \cite{Pervouchine2006}.
\\

\noindent Spike frequency has been shown to affect the skewness of PRCs, using \mbox{type I} integrate-and-fire neurons with adaptation \cite{Gutkin2005}, and to modulate the negative lobe in \mbox{type II} PRCs of conductance based model neurons 
\cite{Fink2011}. Using the aEIF model we have demonstrated that the spike frequency strongly attenuates the effect of either adaptation mechanism on the PRC. At high frequency, unphysiologically large adaptation parameter values are necessary to produce a negative lobe or a significant right-skew in the PRC. This means, for a given degree of adaptation in excitatory neurons, synchronization is possible at frequencies up to a certain value. The stronger the adaptation, the larger this upper frequency limit. 
It has been previously suggested that the degree of adaptation can determine a preferred frequency range for synchronization of excitatory neurons, based on the observation (in vitro and in silico) that the neurons tend to spike in phase with injected currents oscillating at certain frequencies \cite{Fuhrmann2002}. This preferred oscillation frequency increases with increasing degree of SFA. According to our results, at low frequencies synchronization of local circuits through excitatory synapses is possible, provided that the neurons are adapting and delays are short. At higher frequencies, adaptation much less affects the synchronization tendency of excitatory neurons and inhibition may play the dominant role in generating coherent rhythms \cite{Wang1996, Bartos2007}. 
\\

\noindent The adaptation currents $I_m$ and $I_{ahp}$ have previously been found to influence the phase response characteristics of the biophysical Traub neuron model, turning a \mbox{type I} PRC to \mbox{type II} (through $I_m$) and modulating its skew (through $I_{ahp}$) \cite{Ermentrout2001, Ermentrout2012}. We have shown that these changes of the PRC are reflected in the aEIF model by its two adaptation parameters and that in both models (aEIF and Traub) these changes are modulated by the spike frequency. As a consequence, the adaptation induced effects on synchronization of pairs and networks of oscillating neurons are qualitatively similar in both models. Quantitative differences with respect to these effects may well be reduced by fitting the aEIF model parameters to Traub neuron features.
\\

\noindent Our analysis of phase locked states is based on the assumption that synaptic interactions are weak. Experimental work lending support to this assumption has been reviewed in \cite{Hoppensteadt1997, Smeal2010}. Particularly for stellate cells of the entorhinal cortex, synaptic coupling has been found to be weak \cite{Netoff2005a}. Another assumption in this study is that the neurons spike with the same frequency. Considering a pair of neurons spiking at different frequencies, equation \eqref{eq_scalar_phase_ode} needs to be augmented by a scalar $\omega$, which accounts for the constant frequency mismatch between the two neurons \cite{Izhikevich2007}: 
$d\varphi /dt= \omega +H_{21}^d (-\varphi)-H_{12}^d (\varphi)$. In this case, the condition for the existence of phase locked states is $D(\varphi) := H_{21}^d (-\varphi)-H_{12}^d (\varphi)= \omega$. Due to the assumption of weak synaptic strengths however, $\max_\varphi⁡|D(\varphi)|$ must be small, which means that the above condition can only be met if $\omega$ is small. In other words, in the limit of weak coupling phase locking is only possible if the spike frequencies are identical or differ only slightly.
The phase reduction technique considered here, and PRCs in general, are of limited applicability for studying network dynamics in a regime where individual neurons spike at different frequencies, or even irregularly. How adaptation currents affect network synchronization and rhythm in such a regime nevertheless remains an interesting question to be addressed in the future.

\section*{Acknowledgments}

\noindent This work was supported by the DFG Collaborative Research Center SFB910 (JL,MA,KO) and the NSF grant DMS-0908528 (LS). The funders had no role in study design, data collection and analysis, decision to publish, or preparation of the manuscript.

\newpage

\section*{Text S1 - Supplementary Methods}

\subsection*{A) Calculation of the PRC using the adjoint method}

\noindent Let $\mathbf x \in \mathbb R^n$, $\mathbf f: \mathbb R^n \to \mathbb R^n$, and let
$ \mathbf{ \bar x}(t)$ be the $T$-periodic asymptotically stable spiking trajectory as a solution of the system of differential equations 
\begin{equation} \label{eq_uncoupled_ode}
 \frac{d \mathbf x}{dt}  = \mathbf{f}( \mathbf x),
\end{equation}
which describes an uncoupled neuron (cf. Methods). 
In case of the aEIF model, eq.~\eqref{eq_uncoupled_ode} is extended by a reset condition, leading to discontinuities of $ \mathbf{ \bar x}(t)$ at $t \ne kT, \ k \in \mathbb Z$. 
We define the phase $\vartheta \in [0,T)$ of $ \mathbf{ \bar x}(t)$, by a differentiable 1:1-mapping 
$\Theta$ between the points on the periodic spiking trajectory
$\{ \mathbf{\bar x}(t): t \in \mathbb R \}$ and the interval $[0,T)$, 
$\Theta(\mathbf{ \bar x} (\vartheta)) = \vartheta$, where 
$\vartheta = 0$ corresponds to the spike time.
Next, we extend the domain of $\Theta$ to points in the neighborhood of 
$\mathbf{ \bar x}(t)$. Suppose $\mathbf x_0$ is a point on the trajectory $\mathbf{ \bar x}(t)$, $\mathbf y_0$ is a point 
within its domain of attraction, and $\mathbf x (t)$, $\mathbf y (t)$ are the solutions of eq.~\eqref{eq_uncoupled_ode} plus the reset condition in case of the aEIF model with initial conditions $\mathbf x_0$, $\mathbf y_0$. The phase of $\mathbf y_0$, $\Theta(\mathbf y_0)$, is then defined by $\Theta(\mathbf x_0) = \Theta(\mathbf y_0)$ if 
$\lim_{t \to \infty} ||\mathbf x(t) - \mathbf y(t)|| = 0$. \\

\noindent Let $\mathbf {p} \in \mathbb R^n$ be a small perturbation at phase $\vartheta$ which changes the phase of the neuron to $\vartheta_{pert}$. This changes the time of the 
next spike to 
$T_{pert} = \vartheta + T - \vartheta_{pert}$. We then obtain for the PRC
\begin{equation} \label{eq_prc_asymptoticphase}
  \mathrm{PRC}(\vartheta) = 
 T-T_{pert}(\vartheta) = \vartheta_{pert} - \vartheta = 
\Theta(\mathbf{ \bar x}(\vartheta) + \mathbf {p}) - \Theta(\mathbf{ \bar x}(\vartheta))
 = \nabla \Theta(\mathbf{ \bar x}(\vartheta)) ^T \mathbf {p} + O(|| \mathbf {p} ||^2),
\end{equation}
where we have applied Taylor expansion of $\Theta(\mathbf{ \bar x}(\vartheta) + \mathbf {p})$ around $\mathbf{ \bar x}(\vartheta)$.
As $\Theta(\mathbf{ \bar x}(t) + \mathbf {p})$ is rather difficult to calculate, we instead compute $\nabla \Theta(\mathbf{ \bar x}(t))$, 
as explained in the following.\\

\noindent Let $\mathbf{\bar x}(t) + \mathbf z(t)$ be a solution of eq.~\eqref{eq_uncoupled_ode} with initial condition 
$\mathbf{\bar x}(\vartheta) + \mathbf z(\vartheta) = \mathbf{ \bar x } (\vartheta) + \mathbf {p}$ close to the periodic spiking trajectory, i.e. $\mathbf z(t)$ is the deviation from $\mathbf{\bar x}(t)$ for $t \ge \vartheta$. According to the definition of the phase function $\Theta$, the difference 
of the perturbed trajectory's phase and that of the periodic attractor $\mathbf{\bar x}(t)$ is independent of time, that is 
\begin{equation}
\Theta(\mathbf{\bar x}(t) + \mathbf{z}(t)) - \Theta(\mathbf{\bar x}(t)) = c 
  \in \mathbb R \quad \forall t \geq \vartheta,
\end{equation}
which can be rewritten as $\nabla \Theta (\mathbf{\bar x}(t))^T \mathbf z(t) + O(|| \mathbf z(t)||^2) = c$ using Taylor expansion, since $\Theta$ is differentiable.
We neglect terms of second and higher order and define $\mathbf q(t) := \nabla \Theta (\mathbf{\bar x}(t))$ to obtain
\begin{equation} \label{eq_constphasediff}
\mathbf q(t)^T \mathbf z(t) = c \quad \forall t \geq \vartheta.
\end{equation}
For $\vartheta < t < T$, eq.~\eqref{eq_constphasediff} implies 
\begin{equation}
  \frac{d}{dt} \big ( \mathbf q(t)^T \mathbf z(t) \big )
    = \frac{d \mathbf q(t)^T}{dt} \mathbf z(t) + \mathbf q(t)^T 
        D_{\mathbf{x} } \mathbf{f} (\mathbf{\bar x}(t)) \mathbf z(t)
    = \left( \frac{d \mathbf q(t)^T}{dt} + 
        D_{\mathbf{x} } \mathbf{f} (\mathbf{\bar x}(t))^T \mathbf q(t) \right)^T \mathbf z(t)
    = 0,
\end{equation}
where we have used the chain rule and the fact that $\mathbf z(t)$ satisfies the variational 
equation 
\begin{equation} \label{eq_variational}
\frac{d \mathbf z (t)}{dt} = 
D_{\mathbf{x} } \mathbf{f}(\mathbf{\bar x}(t)) \mathbf z (t)
\end{equation}
up to an error of 
$O(|| \mathbf z(t)||^2)$, which can be neglected. Since 
$\mathbf p$ and thus $\mathbf z(t)$ are arbitrary, $\mathbf q(t)$ satisfies the linearized adjoint equation 
\begin{equation} \label{eq_linearadjoint}
  \frac{d \mathbf q(t)^T}{dt} = -D_{\mathbf{x} } \mathbf{f} (\mathbf{\bar x}(t))^T \mathbf q(t).
\end{equation}
In case of the aEIF model we have a discontinuity in $\mathbf{ \bar x}(t)$ for $t=T$. At this point, the displacement $\mathbf z(t)$ of $\mathbf{ \bar x}(t)$ changes discontinuously according to $\mathbf z(T) = A \mathbf z(T^-)$, where 
\begin{equation}
A = \left(\frac{d \bar{V}}{dt}(T^-)\right)^{-1}
\begin{pmatrix}
\displaystyle \frac{d \bar{V}}{dt}(0) & \displaystyle 0 \\
\displaystyle{ \frac{d \bar{w}}{dt}(0) - \frac{d \bar{w}}{dt}(T^-)} & \displaystyle{\frac{d \bar{V}}{dt}(T^-)}
\end{pmatrix}.
 \label{eq_variation_transition} 
\end{equation}
A derivation is provided in \cite{Mueller1995}. The corresponding transition for the adjoint, $\mathbf q(T) = B \mathbf q(T^-)$ can be obtained using eq.~\eqref{eq_constphasediff}, 
\begin{equation} \label{eq_adjoint_jump}
 \mathbf q(T)^T \mathbf z(T) = \mathbf q(T)^T A \mathbf z(T^-) 
   = \big ( B \mathbf q(T^-) \big )^T A \mathbf z(T^-) 
    = \mathbf q(T^-)^T B^T A \mathbf z(T^-) = \mathbf q(T^-)^T \mathbf z(T^-). 
\end{equation}
That is, matrix $B$,
which accounts for the jump of $\mathbf q(t)$ 
is given by $B = A^{-T}$, see e.g. \cite{Samoilenko1995, Akhmet2010}. Note that for a continuous neuron model such as the Traub model, $A=B=I$, where $I$ is the identity matrix.\\

\noindent For $T < \vartheta \leq T + \vartheta$, $\mathbf q(t)$ satisfies the linearized adjoint eq.~\eqref{eq_linearadjoint}, which follows again from eq.~\eqref{eq_constphasediff}.
As $\mathbf q(t)$ is $T$-periodic, it solves eq.~\eqref{eq_linearadjoint} for $t \ne kT, \ k \in \mathbb Z$, and the transition at $t = kT$ is given by $\mathbf q(kT) = B \mathbf q(kT^-)$.
By differentiating $\Theta (\mathbf{\bar x}(\vartheta)) = \vartheta$ with respect to $\vartheta$, we obtain
\begin{align} \label{eq_norm_one_period}
  \mathbf q(\vartheta)^T \frac{d \mathbf{\bar x(\vartheta)}}{d \vartheta} 
    = \mathbf q(\vartheta)^T \mathbf f(\mathbf{\bar x}(\vartheta)) &= 1 
      \quad \forall \vartheta \in (0,T), \: \mathrm{and} \\      
\mathbf q(t)^T \mathbf f(\mathbf{\bar x}(t)) &= 1 
  \quad \forall t \in \mathbb R,  \label{eq_normalization}
\end{align}
using eq.~\eqref{eq_adjoint_jump}, the $T$-periodicity of $\mathbf q(t)$, and the fact that $\mathbf f(\mathbf{\bar x})$ solves the variational 
eq.~\eqref{eq_variational} with transition 
$\mathbf f(\mathbf{\bar x}(kT)) = A \mathbf f(\mathbf{\bar x}(kT^-))$.
We applied eq.~\eqref{eq_normalization} as a normalization condition to determine the appropriate solution of the adjoint system as explained below. \\

\noindent Any $T$-periodic $\mathbf{\tilde q} (t)$ that solves the adjoint system eq.~\eqref{eq_linearadjoint} for $t \ne kT$ and fulfills $\mathbf{ \tilde q}(kT) = B \mathbf{ \tilde q}(kT^-)$ for $t = kT$, can be written as $\mathbf{\tilde q} (t) = \alpha \, \mathbf q(t)$, $\alpha \in \mathbb R$. This follows from the asymptotic stability of $\mathbf{\bar x}(t)$, which implies that $T$-periodic solutions of the variational equation~\eqref{eq_variational} with transition $\mathbf z(kT) = A \mathbf z(kT^-)$ at the discontinuities, are multiples of $\mathbf f( \mathbf{\bar x}(t))$. Thus the space of $T$-periodic solutions of the adjoint system is one-dimensional \cite{Akhmet2010, Ermentrout2010}. The factor $\alpha$ can be determined by requiring that $\mathbf{\tilde q}(t)$ satisfies the normalization condition eq.~\eqref{eq_normalization} for one $t$. This implies that $\mathbf{\tilde q}(t)$ fulfills eq.~\eqref{eq_normalization} for all $t \in \mathbb R$, as can be seen from
\begin{equation} \label{eq_normalization_doesnotchange}
  \frac{d}{dt} \big ( \mathbf{\tilde q}(t)^T \mathbf f( \mathbf{\bar x}(t) )\big ) = 
    -(D_{\mathbf x} \mathbf f(\mathbf{\bar x}(t))^T \mathbf{\tilde q})^T 
      \mathbf f( \mathbf{\bar x}(t) )
      + \mathbf{\tilde q}(t)^T D_{\mathbf x} \mathbf f( \mathbf{\bar x}(t) ) 
      \mathbf f( \mathbf{\bar x}(t) )
    = 0
\end{equation}
for $t \ne kT$ and $\mathbf q(kT)^T \mathbf f( \mathbf{\bar x}(kT) 
  = \mathbf q(kT^-)^T \mathbf f( \mathbf{\bar x}(kT^-)$. \\

\noindent In case of the continuous Traub model, we solve the 
linearized adjoint eq.~\eqref{eq_linearadjoint} numerically backwards in time over several cycles with initial value $\mathbf q(0) = \mathbf f( \mathbf{\bar x}(0))$ to obtain a $T$-periodic solution $\mathbf{\tilde q} (t) = \alpha \, \mathbf q(t)$, and apply the normalization condition eq.~\eqref{eq_normalization} at $t=0$ to fix $\alpha$, i.e.
$\alpha = \mathbf{\tilde q}(0)^T \mathbf f(\mathbf{\bar x}(0))$. For details, see e.g. \cite{Ermentrout2010}. \\

\noindent In case of the aEIF model, $\mathbf q (t)$ is the unique solution to the linearized 
adjoint eq.~\eqref{eq_linearadjoint}, subject to 
the normalization condition 
eq.~\eqref{eq_normalization} at $t=0$, 
\begin{equation}
  \mathbf q(0)^T \mathbf f(\mathbf{\bar x}(0)) = 1, \label{eq_normalization_t0} 
\end{equation}
and the condition 
\begin{equation}
  \mathbf q(0) = B \mathbf q(T^-) \\ \label{eq_adjoint_transition_for_bvp}
\end{equation}
\begin{equation}
  \mathrm{with} \: B = \left(\frac{d \bar{V}}{dt}(0)\right)^{-1}
 \begin{pmatrix}
 \displaystyle \frac{d \bar{V}}{dt}(T^-) & \displaystyle {\frac{d \bar{w}}{dt}(T^-) - \frac{d \bar{w}}{dt}(0)} \\ 
 0 & \displaystyle \frac{d \bar{V}}{dt}(0)
 \end{pmatrix},
\end{equation}
which takes account of the discontinuity and guarantees that $\mathbf q (t)$ is $T$-periodic.
One of the two scalar equations of the latter condition eq.~\eqref{eq_adjoint_transition_for_bvp} can be omitted as explained below. Eq.~\eqref{eq_normalization_t0} implies 
that the normalization condition, eq.~\eqref{eq_normalization}, is satisfied for all $t \in [0,T)$ (cf. eq.~\eqref{eq_normalization_doesnotchange}), including $t = T^-$.
We then obtain 
\begin{equation} \label{eq_adjoint_bvp_derivation1}
    \mathbf q(0)^T \mathbf f(\mathbf{\bar x}(0)) = 
    \mathbf q(T^-)^T \mathbf f(\mathbf{\bar x}(T^-))
\end{equation}
\begin{align}
  & \Longleftrightarrow q^V(0) \frac{d \bar V}{dt} (0) + q^w(0) \frac{d \bar w}{dt} (0) = 
    q^V(T^-) \frac{d \bar V}{dt} (T^-) + q^w(T^-) \frac{d \bar w}{dt} (T^-)
    \label{eq_adjoint_bvp_derivation2} \\ 
  & \Longleftrightarrow q^V(0) \frac{d \bar V}{dt} (0) = \frac{d \bar V}{dt} 
    (T^-) q^V(T^-) + \left ( \frac{d \bar w}{dt} (T^-) - \frac{d \bar w}{dt} (0)  
    \right ) q^w(T^-), \label{eq_adjoint_bvp_derivation3}
\end{align}
where eq.~\eqref{eq_adjoint_bvp_derivation3} is the first scalar equation of eq.~\eqref{eq_adjoint_transition_for_bvp}
multiplied with $\tfrac{d \bar V}{dt} (0)$. In eq.~\eqref{eq_adjoint_bvp_derivation3} we have used the second scalar 
equation of eq.~\eqref{eq_adjoint_transition_for_bvp}, 
\begin{equation}
  q^w(0) = q^w(T^-). \label{eq_adjoint_transition_for_bvp_secondscalar}
\end{equation}
It follows that if eqs.~\eqref{eq_normalization_t0} and 
\eqref{eq_adjoint_transition_for_bvp_secondscalar} are satisfied, eq.~\eqref{eq_adjoint_bvp_derivation3} and thus the first scalar 
equation of eq.~\eqref{eq_adjoint_transition_for_bvp} hold as well. 
It is therefore sufficient to solve eq. \eqref{eq_linearadjoint} for $t \in (0,T)$ using 
conditions \eqref{eq_normalization_t0} and 
\eqref{eq_adjoint_transition_for_bvp_secondscalar}. This is equivalent the 
boundary value problem, eqs.~(17)--(20) from the Methods section of the main paper.\\

\noindent As the synaptic current only perturbs the membrane potential, the perturbation $\mathbf {p} = (p_1,0,\dots,0)^T$ considered here is nonzero only in the first component. Thus, the PRC reduces to $q^V(\vartheta) \, p_1$, where $q^V$ denotes the first component of $\mathbf q$. Since $p_1$ is only a scaling factor, 
in this study we identify $q^V$ with the PRC.
\\

\subsection*{B) Phase reduction}

\noindent In the following we describe how the full network model eq.~(10) is reduced to a lower dimensional network model where each neuron is represented by its phase $\vartheta_i$. This phase reduction requires weak coupling between each pair of neurons which we emphasize by rewriting $I_{syn}(V_i,V_j) = \varepsilon \tilde I_{syn}(V_i,V_j)$, where $I_{syn}$ is the synaptic current introduced in eq.~(11) and
$\varepsilon > 0$ is small (due to small conductance $g_{ij}$).
By applying a change of variables $\vartheta_i := \Theta(\mathbf x_i)$ in eq.~(10), with phase function $\Theta$ as defined in the previous section, the network equation for neuron $i$ becomes
\begin{align}
 \frac{d \vartheta_i}{dt} & = \frac{d \Theta(\mathbf x_i)}{dt}
   = \nabla \Theta(\mathbf x_i) ^T \frac{d \mathbf x_i}{dt} \label{eq_phasenet1} \\
   & = \nabla \Theta(\mathbf x_i) ^T \mathbf{f}(\mathbf x_i) + \nabla \Theta(\mathbf x_i) 
  ^T \sum_{j=1}^N \mathbf h_{ij}(\mathbf x_i, \mathbf x_j) \\ 
   & = 1 + \varepsilon \frac{\partial \Theta(\mathbf x_i)}{\partial x_1}  
          \sum_{j=1}^N \tilde I_{syn}(\mathbf x_i, \mathbf x_j). \label{eq_phasenet2}
\end{align}
In eqs.~\eqref{eq_phasenet1}--\eqref{eq_phasenet2} we have used the chain rule, the relation $ \nabla \Theta(\mathbf x_i) ^T \mathbf{f}(\mathbf x_i) = 1$ which is evident when considering the uncoupled system, and the fact that the coupling function 
$\mathbf h_{ij}(\mathbf x_i, \mathbf x_j)$ is nonzero only in the first component where it consists of $\varepsilon \tilde I_{syn}(\mathbf x_i, \mathbf x_j)$.
Next, to get rid of the state variables $\mathbf x_i$ in eq.~\eqref{eq_phasenet2}, we first approximate 
$\mathbf x_i$ using the periodic spiking trajectories parametrized by the phase $\mathbf{ \bar x}_i(\vartheta_i)$. 
This approximation causes an error of $O(\varepsilon)$, which becomes $O(\varepsilon^2)$ due to the factor $\varepsilon$,
\begin{equation}
 \frac{d \vartheta_i}{dt}  = 1 + \varepsilon \frac{\partial \Theta(\mathbf{\bar x}_i(\vartheta_i))}{\partial x_1}
 \sum_{j=1}^N \tilde I_{syn}(\mathbf{\bar x}_i(\vartheta_i), \mathbf{\bar x}_j(\vartheta_j)) + O(\varepsilon^2).
\end{equation}
We neglect second order terms in $\varepsilon$ and apply another change of variables  $\psi_i := \vartheta_i - t$,
\begin{equation}
  \frac{d \psi_i}{dt} = \varepsilon  \frac{\partial \Theta(\mathbf{\bar x}_i(t + \psi_i))}{\partial x_1} 
  \sum_{j=1}^N \tilde I_{syn}(\mathbf{\bar x}_i(t + \psi_i), \mathbf{\bar x}_j(t + \psi_j)),
\end{equation}
to obtain an equation to which we can apply the method of averaging, see e.g. \cite{Hoppensteadt1997}, that leads to
\begin{align}
 \frac{d \bar \psi_i}{dt} &= \varepsilon \frac{1}{T} \int_0^T \frac{\partial 
 \Theta(\mathbf{\bar x}_i(s + \bar \psi_i))}{\partial x_1} 
 \sum_{j=1}^N \tilde I_{syn}(\mathbf{\bar x}_i(s + 
      \bar \psi_i), \mathbf{\bar x}_j(s + \bar \psi_j)) ds \\
  &=  \varepsilon \sum_{j=1}^N \frac{1}{T} \int_0^T \frac{\partial 
   \Theta(\mathbf{\bar x}_i(s))}{\partial x_1} 
   \tilde I_{syn}(\mathbf{\bar x}_i(s), \mathbf{\bar x}_j(s + \bar \psi_j - \bar \psi_i)) ds, \label{eq_averaged_phase}    
\end{align}
where we have used in eq.~\eqref{eq_averaged_phase} that the spiking trajectories are $T$-periodic. 
Changing the variables one more time $\bar \vartheta_i := t + \bar \psi_i$ we arrive at
\begin{equation} 
   \frac{d \bar \vartheta_i}{dt} = 1 + \varepsilon \sum_{j=1}^N \frac{1}{T} \int_0^T \frac{\partial 
   \Theta(\mathbf{\bar x}_i(s))}{\partial x_1} 
   \tilde I_{syn}(\mathbf{\bar x}_i(s), \mathbf{\bar x}_j(s + \bar \vartheta_j - \bar \vartheta_i)) ds,
\end{equation}   
which is identical to eq.~(21), recognizing that $\partial \Theta(\mathbf{\bar x}_i(s)) / \partial x_1 = q_i^V (s)$ and $\varepsilon \tilde I_{syn}(V_i,V_j) = I_{syn}(V_i,V_j)$. Note that the phases $\vartheta_i$ in
eqs.~(21) and (22) are averaged phases $\bar \vartheta_i$.
\\

\subsection*{C) Relation between the PRC and the slope of the interaction function}

Let the interaction function $H^d(\varphi)$ be 
\begin{equation}
H^d(\varphi) = \frac{1}{T} \int_0^T 
  q^V (\vartheta) \, g \, s(\vartheta + \varphi -d )
        (E_{syn}-\bar V(\vartheta) ) \, d \vartheta ,  \label{H_function1}
\end{equation}
according to
eqs.~(11),~(12),~(21),~(22) of the main paper, for a pair of identical neurons with symmetric synaptic strengths $g$ and equal conduction delays $d$. Phase locked states, i.e. the roots $\varphi_0$ of $H_{\Delta}(\varphi) = H^{d}(-\varphi)-H^{d}(\varphi)$ (cf. eq.~(23)), are stable if $\mathrm{lim}_{\varepsilon \searrow 0} \, d H^{d}(\varphi_0 - \varepsilon) / d \varphi > 0$ and $\mathrm{lim}_{\varepsilon \searrow 0} \, d H^{d}(\varphi_0 + \varepsilon) / d \varphi > 0$ (see the section Results). The left and right sided limits differ for $\varphi_0 = d$ in case of aEIF neurons, and are equal otherwise.
Below, we explain how the sign of these limits are determined by the PRC ($q^V$).
\\ \\
Consider $\varphi_0 \neq d$. Then 
\begin{equation}  \label{eq_limits}
  \mathrm{lim}_{\varepsilon \searrow 0} 
    \frac{d}{d \varphi} H^d(\varphi_0 - \varepsilon)
  = \mathrm{lim}_{\varepsilon \searrow 0} 
    \frac{d}{d \varphi} H^d(\varphi_0 + \varepsilon)
  = \frac{d}{d \varphi} H^d(\varphi_0) \:\: \mathrm{and}
\end{equation}
\begin{align}
\mathrm{sgn} \, \frac{d}{d \varphi} H^d(\varphi_0) &= 
\mathrm{sgn} \, \frac{d}{d \varphi} \int_0^T
  q^V (\vartheta) \, s(\vartheta + \varphi_0 - d)
        (E_{syn}- \bar V(\vartheta) ) \, d \vartheta  \label{eq_sgn1} \\
& = \pm \, \mathrm{sgn} \, \frac{d}{d \varphi} 
	\int_{\varphi_0 - d}^{T + \varphi_0 - d}	
   q^V (t - \varphi_0 + d) \, s(t) \, dt \label{eq_sgn2} \\
& = \mp \, \mathrm{sgn} \left( \int_{\varphi_0 - d}^{T + \varphi_0 - d}	
  \frac{d}{d \varphi} q^V (t - \varphi_0 + d) \, s(t) \, dt 
  \right.\nonumber \\
& \qquad \qquad \left.\vphantom{\int_t} + \left( q^V (0) - q^V (T^-) \right) \, s(\varphi_0 - d) 
  \right). \label{eq_sgn3}    
\end{align} 
In eq.~\eqref{eq_sgn2} we have replace the factor $(E_{syn}-\bar V(\vartheta))$ by $+1$ for excitatory synapses and $-1$ for inhibitory synapses before applying a change of variables $t = \vartheta + \varphi_0 - d$. The replacement is justified, as in the former case $(E_{syn}-\bar V(\vartheta))$ is positive over the period $T$ except for a very brief interval during the spike and in the latter case $(E_{syn}-\bar V(\vartheta))$ is negative except for a possible transient interval of hyperpolarization (below $E_{syn}$) just after the spike. 
In eq.~\eqref{eq_sgn3} we have used the Leibniz integral rule. Note that in 
eqs.~\eqref{eq_sgn1}--\eqref{eq_sgn3} the integral are over half-open intervals and that $q^V(\vartheta)$ is differentiable for $\vartheta \in (0, T)$. Since $s(t)$ is a continuous $T$-periodic function,
eq.~\eqref{eq_sgn3} also holds for $\varphi_0 = d$, and therefore the sign of the limits in eq.~\eqref{eq_limits} is given by eq.~\eqref{eq_sgn3}.
\\ \\
We assume that $s(t)$ has finite rise and decay times, such that at some time point $t = \varepsilon_{syn} > 0$, $s(t)$ has decayed to a sufficiently small value. It becomes evident that if the sign of $q^V(\vartheta)$ remains unchanged over the interval $\vartheta \in (d - \varphi_0, d - \varphi_0 + \varepsilon_{syn})$ and, in case the interval contains 0 or $T$, 
$\mathrm{sgn} (q^V(\vartheta)) = \mathrm{sgn} (q^V (0) - q^V (T^-))$, then the sign in eq.~\eqref{eq_sgn3} and thus the stability of $\varphi_0$ is determined.
\\

\newpage


%

\newpage

\section*{Figures}

\newpage

\newpage

\newpage

\end{document}